\documentclass[prd,aps,nofootinbib,showpacs,superscriptaddress,amsmath,amssymb,aps]{revtex4-1}
\usepackage[bottom]{footmisc}
\usepackage{graphicx}
\usepackage{dcolumn}
\usepackage{bm}
\usepackage{natbib,times}
\usepackage[normalem]{ulem}
\usepackage{xcolor}
\usepackage{xspace}

\def\fun#1#2{\lower3.6pt\vvbox{\baselineskip0pt\lineskip.9pt
  \ialign{$\mathsurround=0pt#1\hfil##\hfil$\crcr#2\crcr\sim\crcr}}}
\def\simgt{\mathrel{\lower0.6ex\hbox{$\buildrel {\textstyle >}
 \over {\scriptstyle \sim}$}}}
\def\simlt{\mathrel{\lower0.6ex\hbox{$\buildrel {\textstyle <}
 \over {\scriptstyle \sim}$}}}

\def\bea{\begin{eqnarray}}
\def\eea{\end{eqnarray}}
\def\be{\begin{equation}}
\def\ee{\end{equation}}
\def\bes{\begin{split}}
\def\ees{\end{split}}
\def\ba{\begin{eqnarray}}
\def\ea{\end{eqnarray}}

\def\m{{\rm m}}
\def\({\Big(}
\def\){\Big)}

\def\a{\alpha}
\def\b{\beta}

\def\m{\mu}
\def\n{\nu}

\newcommand\bw{\begin{widetext}}
\newcommand\ew{\end{widetext}}

\newtheorem{definition}{Definition}
\newtheorem{lemma}{Lemma}

\newcommand{\proofof}[1]{\noindent {\bf Proof of #1. }}
\newcommand{\qed}{\hfill $\fbox{\hspace{0.3mm}}$ \vspace{.3cm}} 

\newcommand{\Real}{\mathbb{R}}
\newcommand{\ve}[1]{\underline{#1}}

\begin{document}

\title{Well-posed Cauchy formulation for Einstein-\ae ther theory}

\author{Olivier Sarbach}
\affiliation{Instituto de F\'isica y Matem\'aticas,
Universidad Michoacana de San Nicol\'as de Hidalgo,
Edificio C-3, Ciudad Universitaria, 58040 Morelia, Michoac\'an, M\'exico}
\email{sarbach@ifm.umich.mx}

\author{Enrico Barausse}
\affiliation{Institut d'Astrophysique de Paris, CNRS \& Sorbonne
 Universit\'es, UMR 7095, 98 bis bd Arago, 75014 Paris, France}
\affiliation{SISSA, Via Bonomea 265, 34136 Trieste, Italy and INFN Sezione di Trieste}
\affiliation{IFPU - Institute for Fundamental Physics of the Universe, Via Beirut 2, 34014 Trieste, Italy}
\email{barausse@sissa.it}

\author{Jorge A. Preciado-L\'opez}
\affiliation{Perimeter Institute for Theoretical Physics, 31 Caroline Street North, Waterloo, ON, N2L 2Y5, Canada}
\email{jpreciado@perimeterinstitute.ca}

\date{\today}

\begin{abstract}
We study the well-posedness of the initial value (Cauchy) problem of vacuum 
Einstein-\ae ther theory. The latter is a Lorentz-violating gravitational theory consisting
of General Relativity with a dynamical timelike ``\ae ther'' vector field, which
selects a ``preferred time'' direction at each spacetime event. The Einstein-\ae ther action
is quadratic in the \ae ther, and thus yields second order field
equations for the metric and the \ae ther. However, the well-posedness of the Cauchy problem is not easy to prove away from
the simple case of perturbations over flat space. This is particularly problematic because well-posedness is a 
necessary requirement to ensure stability of numerical evolutions of the initial value problem.
Here, we employ a first-order formulation of Einstein-\ae ther theory in terms of projections on a tetrad frame. 
We show that under suitable conditions on the coupling constants of the theory, the resulting evolution equations can be cast into strongly or even symmetric hyperbolic form, and therefore they define a well-posed Cauchy problem.
\end{abstract}

\maketitle

\section{\label{sec:level1}Introduction} 

Lorentz symmetry has long been one of the cornerstones of theoretical physics, and has been tested to high precision in a variety of experiments.
The Standard Model of particle physics is obviously Lorentz invariant, but parametrized formalisms such as
the Standard Model Extension~\cite{Colladay:1998fq, Kostelecky:1998id,Kostelecky:1999rh} have been introduced and used to place very strong
constraints on Lorentz violations (LVs) in matter~\cite{Kostelecky:2003fs,Kostelecky:2008ts,Mattingly:2005re,Jacobson:2005bg} and in 
the matter-gravity sector~\cite{Kostelecky:2010ze}. As for the purely gravitational sector, tests of LVs have historically been less compelling, partly
because of the absence of a parametrized formalism applicable to strong gravitational fields. Indeed, theory independent tests of Lorentz symmetry 
can be performed in the solar system~\cite{Will:2014kxa} and in binary/isolated pulsars~\cite{1987ApJ...320..871N,Damour:1992ah,Shao:2013wga,Shao:2012eg} 
at first Post-Newtonian (PN) order\footnote{The PN formalism~\cite{Blanchet:2013haa} is an expansion
of the dynamics in powers of $v/c$, $v$ being the characteristic velocity of the
system. Terms of
order $(v/c)^{2 n}$ relative to the leading one are referred to as of ``nPN'' order.} in the conservative dynamics,
and more recently with the propagation of gravitational waves (by parameterizing their dispersion relation~\cite{Monitor:2017mdv}), but it
is very difficult to extend these tests to the highly dynamical, relativistic and strong-field regime relevant e.g. to sources for Advanced LIGO and Virgo.
To understand the effect of LVs in these systems, it is much more fruitful to consider specific gravitational theories explicitly breaking Lorentz symmetry.

General Relativity (GR) is of course Lorentz (and diffeomorphism) invariant, but the gravitational theory can be made Lorentz violating
by introducing additional Lorentz violating gravitational degrees of freedom besides the spin-2 polarizations of GR. 
Focusing, for concreteness, on boost violations rather than on general LVs, these extra graviton polarizations
can be for instance a vector field $u^\mu$ -- constrained to be timelike at the level of the action,
so as to represent a ``preferred time'' direction at each spacetime event -- or a scalar field
-- again constrained to have a timelike gradient $u^\mu$ at the level of the action, so as to represent a ``preferred foliation'' of the spacetime.
By writing the most generic gravitational action (to quadratic order in the gradient of $u^\mu$) for these vector or scalar degrees of freedom,
one obtains respectively Einstein-\ae ther theory~\cite{Jacobson:2000xp} and khronometric gravity~\cite{Blas:2009qj,Horava:2009uw,Jacobson:2010mx}. These are indeed the most generic boost violating extensions of GR at low energies, and
have been extensively used as a framework to place bounds on LVs in purely gravitational experiments, e.g. from solar system tests~\cite{Foster:2005dk,Jacobson:2008aj,Blas:2011zd,Bonetti:2015oda}; from the coincident 
gravitational and electromagnetic detection of GW170817 and GRB 170817A~\cite{Monitor:2017mdv,Gumrukcuoglu:2017ijh}; from binary and isolated pulsars~\cite{Foster:2007gr,Yagi:2013ava,Yagi:2013qpa}; and from 
isolated~\cite{Eling:2006ec,Barausse:2011pu,Blas:2011ni} and binary black holes~\cite{Ramos:2018oku}. Similarly, one can in principle extend GR by breaking 
invariance under spatial diffeomorphisms (i.e. spatial rotations), see e.g.~\cite{Dubovsky:2004sg,Blas:2009my}. 
In this paper, however, we will focus on boost-violating theories, and more specifically on Einstein-\ae ther theory.

It should be stressed that boost-violating theories are not only interesting from a phenomenological point of view (as strawmen to test LVs in gravity), but also
from a more fundamental perspective. Indeed, by violating boost symmetry one can introduce anisotropies (Lifschitz scaling) between the time and spatial coordinates,
which result in a better ultraviolet (UV) behavior of the theory. Indeed, khronometric theory turns out to coincide with the low-energy limit of
Ho\v rava gravity~\cite{Horava:2009uw}, which, unlike GR, is power counting~\cite{Horava:2009uw} and also perturbatively renormalizable~\cite{Barvinsky:2015kil},
thanks exactly to the presence of the aforementioned anisotropic scaling between space and time. While it is yet unclear whether
Ho\v rava gravity can be a fully viable theory of quantum gravity when matter is included -- mainly
because one needs to suppress (e.g. via renormalization group running~\cite{Chadha:1982qq,Bednik:2013nxa,Barvinsky:2017kob}\footnote{Note however that Ref.~\cite{Knorr:2018fdu} finds that the renormalization group flow may not approach GR at low energies.},
supersymmetry~\cite{GrootNibbelink:2004za} or  a large energy scale~\cite{Pospelov:2010mp}) the percolation of large LVs from gravity to the matter sector,
the improved UV behavior makes boost-violating gravitational theories particularly attractive.

A conspicuous practical problem with studying gravitational wave emission from systems of two compact objects (neutron stars or black holes) in theories
extending GR is (in general) the absence of results on the stability of the initial value (Cauchy) problem.  
In GR, the Cauchy problem can be put in a ``well-posed'' form, i.e. such that for given initial data there exists a unique time evolution which depends continuously on the initial data (see e.g. Refs.~\cite{Wald:1984rg,Sarbach:2012pr} for reviews). This
property is clearly crucial to integrate systems of compact objects on a computer (currently the only way we have to
study rigorously their merger), as it prevents numerical errors to grow unbounded. 
Very few results for the well-posedness of the Cauchy problem exist beyond GR, with the exception of scalar tensor theories of
the Fierz-Jordan-Brans-Dicke type~\cite{Fierz:1956zz,Jordan:1959eg,Brans:1961sx} (see e.g. Refs.~\cite{Salgado:2005hx,Barausse:2012da}). Attempts have been made to enforce stability of the initial value problem via reduction of
order techniques~\cite{Okounkova:2017yby} or by ``smoothing'' higher derivatives (e.g. using techniques from relativistic hydrodynamic)~\cite{Cayuso:2017iqc,Allwright:2018rut}, but it is unclear how
these procedures impact the results of numerical simulations. 

From this point of view, boost-violating gravity is a perfect case study to assess the stability (or lack thereof) in theories of gravity
extending GR. As mentioned, both khronometric and Einstein-\ae ther present additional graviton polarizations besides those
of GR, namely a spin-0 graviton in the former~\cite{Blas:2011zd} and a spin-0 graviton and two spin-1 polarizations in the latter~\cite{Jacobson:2004ts}. In both theories,
these extra gravitons satisfy wave equations in flat space, for generic values of the theory parameters, but it is unclear if
similar results apply on curved backgrounds. Nevertheless, the flat space results provide hope that the Cauchy problem may be well posed
on generic backgrounds.

In the following we will focus on Einstein-\ae ther theory, whose dynamics is richer and more complicated than khronometric theory because of the presence of spin-1 degrees of freedom. In section~\ref{Sec:AE} we briefly review the action and field equations of the theory, as well as the experimental constraints on the coupling constants. In section~\ref{toy} we present a toy problem highlighting the idea that inspired us to use a frame formulation of Einstein-\ae ther theory, which we present in section~\ref{frame_formulation}. We proceed to show in section~\ref{strong_hyper} that under suitable conditions on the coupling constants, the field equations in this frame formulation of the theory are strongly hyperbolic. Next, in section~\ref{sym_hyper} we derive a three-parameter subfamily of this formulation which satisfies the stronger requirement of symmetric hyperbolicity and thus yields a well-posed Cauchy problem. We briefly summarize our conclusions in section~\ref{conclusions} and include technical details in appendices.

Throughout this paper, we will use the $(-+++)$ metric signature and set the speed of light $c=1$.
We will also introduce the following combinations of the coupling constants $c_i$ (with $i=1,2,3,4$) of Einstein-\ae ther theory: $c_{ij}\equiv c_i+c_j$
and $c_{ijk} \equiv c_i+c_j+c_k$.

\section{Einstein-\ae ther theory: action and field equations}
\label{Sec:AE}

Einstein-\ae ther theory breaks boost invariance explicitly in the gravitational sector by
introducing a preferred time  ``direction'' at each spacetime event, via a timelike 
``\ae ther'' vector field $u^\alpha$ with unit norm $g_{\alpha\beta} u^\alpha u^\beta=-1$.
The most generic covariant modification to the GR action that is quadratic in the \ae ther
is then given by~\cite{Jacobson:2000xp}
\begin{equation}
S = \frac{1}{16\pi G_{\text{\ae}}} \int 
\left[ R - M^{\alpha\beta}{}_{\mu\nu}(\nabla_\alpha u^\mu)(\nabla_\beta u^\nu) 
 +\lambda (g_{\alpha\beta} u^\alpha u^\beta + 1) \right]
\sqrt{-g} d^4 x+ S_m\left[\psi,g_{\alpha\beta}\right],
\label{Eq:AEAction}
\end{equation}
where
\begin{equation}
M^{\alpha\beta}{}_{\mu\nu} = c_{1} g^{\alpha\beta} g_{\mu\nu} 
 + c_{2}\delta^\alpha{}_\mu\delta^\beta{}_\nu
 + c_{3}\delta^\alpha{}_\nu\delta^{\beta}{}_\mu 
 - c_{4}u^\alpha u^\beta g_{\mu\nu},
\label{Eq:MDef}
\end{equation}
 $c_1$, $c_2$, $c_3$, $c_4$ denote four dimensionless coupling constants, $R$ is the Ricci scalar, $\lambda$ is a Lagrange multiplier that enforces the unit norm constraint on the \ae ther field, and $G_{\text{\ae}}$ is the bare gravitational constant, related to the value $G_N$ measured in the solar system by $G_N = G_{\text{\ae}}/[1-c_{14}/2]$~\cite{Carroll:2004ai}. The matter fields, collectively denoted by $\psi$ and appearing in the matter action $S_m$, are then supposed to couple
only to the metric at tree level, so as to avoid the appearance of unwanted ``fifth forces'' and to pass existing particle physics tests of Lorentz invariance (c.f. the discussion in the Introduction). In the following, however, we will focus on the character of the Cauchy problem in vacuum, and we will therefore set $S_m=0$.

Varying the vacuum action with respect to $g^{\alpha\beta}$, $u^\beta$ and $\lambda$ yields the field equations
\begin{eqnarray}
G_{\alpha\beta} &=& T^{\text{\ae}}_{\alpha\beta},
\label{Eq:EoMg}\\
\nabla_\alpha J^\alpha{}_\beta + c_4 a_\alpha\nabla_\beta u^\alpha &=& -\lambda u_\beta,
\label{Eq:EoMu}\\
u^\alpha u_\alpha &=& -1,
\label{Eq:EoMlambda}
\end{eqnarray}
where $G_{\alpha\beta}$ denotes the Einstein tensor, $T^{\text{\ae}}_{\alpha\beta}$ is the \ae ther stress-energy tensor explicitly given by
\begin{eqnarray}
T^{\text{\ae}}_{\alpha\beta} 
 &=& \nabla_{\mu} \left( J^{\mu}{}_{(\alpha} u_{\beta)} - J_{(\alpha}{}^{\mu} u_{\beta)} 
 + J_{(\alpha\beta)} u^\mu  \right)   \nonumber\\
 &+& c_1\left[ (\nabla_\alpha u^\mu)(\nabla_{\beta} u_\mu) 
 - (\nabla^\mu u_\alpha) (\nabla_\mu u_\beta)  \right]
 - \frac{1}{2} g_{\alpha\beta} J^\mu{}_\nu\nabla_\mu u^\nu
 + c_4 a_\alpha a_\beta + \lambda u_\alpha u_\beta,
\label{Eq:TaeDef}
\end{eqnarray}
and where we have introduced
\begin{eqnarray}
J^\alpha{}_\mu &\equiv& M^{\alpha\beta}{}_{\mu\nu} \nabla_{\beta}u^\nu, \\
a_\alpha &\equiv& u^\mu\nabla_\mu u_\alpha. 
\end{eqnarray}
Note that the Lagrange multiplier can be eliminated from the system by projecting
Eq.~\eqref{Eq:EoMu} on the space orthogonal to the \ae ther, or equivalently by contracting the same equation
with $u^\beta$ (thus solving for $\lambda$).

Expanding the field equations over flat space reveals that the theory has additional degrees of freedom compared to GR. Indeed,
the theory presents two spin-2 polarizations, as in GR, but also one spin-0 and two spin-1 polarizations. The (squared) propagation speeds
on flat space are respectively given by~\cite{Jacobson:2004ts}
\begin{gather}
s_2^2=\frac{1}{1-c_{13}}\, ,
\label{s2}\\
s_1^2=\frac{2c_1- c_1^2+
c_3^2}{2c_{14}(1-c_{13})} \, ,
\label{s1}\\
s_0^2=\frac{c_{123}(2-c_{14})}{c_{14}(1-c_{13})(2+c_{13}+3c_2)}\, .
\label{s0}
\end{gather}
Stability at the classical and quantum levels requires $s_2^2$, $s_1^2$ and $s_0^2$ to be positive~\cite{Jacobson:2004ts,Garfinkle:2011iw}. 
Moreover, ultrahigh energy cosmic ray observations
require $s_i^2\gtrsim1-{\cal O}(10^{-15})$ (with $i=0,1,2$), to prevent cosmic rays from losing energy into gravitational modes via a Cherenkov-like cascade~\cite{Elliott:2005va}. More recently, the coincident gravitational and
electromagnetic detection of GW170817 and GRB 170817A has constrained $-3\times10^{-15} <s_2-1< 7\times10^{-16}$~\cite{Monitor:2017mdv}. 

Additional bounds come from the requirement that the theory should agree with solar system experiments. At 1PN order and for weakly gravitating sources 
such as those encountered in the solar system, the theory only deviates from GR through the preferred frame parameters $\alpha_1$ and $\alpha_2$ appearing in the
parametrized PN expansion~\cite{Will:2014kxa}. Those parameters are related to the coupling constants by~\cite{Foster:2005dk}
\begin{gather}
    \alpha_1= \frac{-8(c_3^2 + c_1c_4)}{2c_1 - c_1^2+c_3^2}\, ,
\label{eq:alpha1}\\
    \alpha_2=\frac{\a_1}{2}
    -\frac{(c_1+2c_3-c_4)(2c_1+3c_2+c_3+c_4)}{c_{123}(2-c_{14})} \, .
\label{eq:alpha2}
\end{gather}
Solar systems tests require $|\alpha_1|\lesssim 10^{-4}$ and $|\alpha_2|\lesssim10^{-7}$~\cite{Will:2014kxa}. \textit{Saturating} these constraints (i.e. assuming in particular that $|\alpha_1|\lesssim 10^{-4}$ but \textit{not} $|\alpha_1|\ll 10^{-4}$), together with the aforementioned constraints on the propagation speeds, yields $c_1\approx -c_3+{\cal O}(10^{-15})$, $c_4\approx c_3+{\cal O}(10^{-4})$, and $c_2\approx (c_4-c_3) [1+{\cal O}(10^{-3})]$. Therefore, to within an accuracy of $10^{-4}$ or better in the coupling parameters, Einstein-\ae ther theory possesses a one-dimensional viable parameter space, i.e. one has $c_1+c_3\approx0$, $c_4-c_3\approx c_2\approx0$, while $c_1-c_3$ is essentially unconstrained. This latter combination corresponds to the coefficient multiplying the \ae ther vorticity in the action, as shown explicitly in Ref.~\cite{Jacobson:2013xta}. In particular, it can be easily shown that one can send $|c_1-c_3|\to\infty$ or $|c_1-c_3|\to0$, while passing all aforementioned experimental bounds.

Note that gravitational wave generation (e.g. in binary pulsars/black holes) is not expected to further constrain the theory at low PN orders, at least 
in the limits $|c_1-c_3|\to\infty$ and $|c_1-c_3|\to0$. Indeed, in both limits gravitational wave emission should approach the 
GR predictions. This happens, in the latter case, because all the coupling parameters $c_i$ go to zero (to within ${\cal O}(10^{-4})$ or better),
hence the action of the theory approaches that of GR. In the former case, instead, one
can show~\cite{Jacobson:2013xta,Barausse:2015frm} that the theory's {\it solutions} converge to
those of a khronometric theory with coupling parameters of ${\cal O}(10^{-4})$ or smaller, hence 
deviations from GR in the solutions should be small. We leave the task of deriving 
detailed predictions for gravitational wave emission (especially away from these two limits) to
future work, as for the present one it is sufficient to show that the theory still has a viable parameter space.\footnote{Note that
constraints from binary pulsars were derived in Ref.~\cite{Yagi:2013ava} in a different portion of the parameter space
than the one currently favored after the GW170817 detection. Indeed, neutron star sensitivities, which are a crucial ingredient to
constrain gravitational theories with pulsar data, where computed under the assumption that $\alpha_1=\alpha_2=0$. That choice, together
with the requirement that the graviton propagation speeds be $\gtrsim 1$ to avoid vacuum Cherenkov radiation, excluded the 
parameter space ``line'' with $c_1+c_3\approx0$, $c_4-c_3\approx c_2\approx0$ and variable $c_1-c_3$ discussed above.}

We note in passing that another viable portion of the parameter space may be obtained by requiring that $|\alpha_1|$ be much smaller than its bound, so that the bound on $\alpha_2$ is also satisfied automatically (since $\alpha_2\propto{\a_1}$ if $c_1+c_3=0$, as required by GW170817). Indeed, much of the literature about Einstein-\ae ther theory prior to GW170817 set $\alpha_1$ and $\alpha_2$ exactly to zero. Such a choice, combined with the bound from GW170817, yields $c_4=c_3$ and thus a two-dimensional parameter  space $(c_2,c_1-c_3)$. However, both the spin-0 and spin-1 propagation speeds diverge in this limit. Of course, requiring $0\neq|c_3-c_4|\lesssim 10^{-7}$ provides in principle a viable two-dimensional parameter  space $(c_2,c_1-c_3)$ (with the only further requirement that $|c_2|\lesssim 0.1$ to pass Big Bang Nucleosynthesis bounds~\cite{Carroll:2004ai}) and large but finite speeds. These large speeds also make the spin-0 and spin-1 fields non-dynamical and therefore suppress the deviations away from GR in gravitational wave data (e.g. binary pulsars)~\cite{Yagi:2013ava,Ramos:2018oku}. While such a choice is less generic than that of saturating the bound on $|\alpha_1|$, it is in principle a possibility. 

Having considered all presently known experimental constraints on the theory, here we will focus on a new class of theoretical bounds that have never been considered thus far. In more detail, in this work we aim to investigate whether  the system given by Eqs.~(\ref{Eq:EoMg},\ref{Eq:EoMu},\ref{Eq:EoMlambda}) can provide a well-posed Cauchy problem. Indeed, while perturbations over flat space do produce a strongly hyperbolic system (because they can be recast as wave equations for the spin-2, spin-1 and spin-0 modes), it is unclear if the same result can be obtained for the full system. The main difficulty consists in the second derivatives of the metric fields $g_{\alpha\beta}$, which appear in the effective stress-energy tensor $T^{\text{\ae}}_{\alpha\beta}$ through the second covariant derivatives of $u^\alpha$.

\section{Toy model example}
\label{toy}

To gain some insight into the well-posedness of the initial value problem in Einstein-\ae ther theory, we start with a simpler toy theory with a somewhat similar structure. To this purpose we consider a $U(1)$ gauge field $A_\mu$ on flat spacetime which is coupled to a complex massless scalar field $\Phi$ of charge $q\neq 0$ subject to the constraint $|\Phi| = 1$. The corresponding action is
\begin{equation}
S_{toy} = \int \left[ -\frac{1}{4} F^{\mu\nu} F_{\mu\nu} 
 - \frac{1}{2} (D^\mu\Phi)^*(D_\mu\Phi) - \frac{\lambda}{2}(|\Phi|^2 - 1) \right] d^4 x,
\label{Eq:ToyAction}
\end{equation}
where $F_{\mu\nu} = \partial_\mu A_\nu - \partial_\nu A_\mu$ is the Faraday tensor, $D_\mu = \partial_\mu + i q A_\mu$ is the covariant derivative operator, and $\lambda$ is a real Lagrange multiplier. The equations of motion are
\begin{eqnarray}
\partial_\nu F^{\mu\nu} &=& i\frac{q}{2}\left[ \Phi^* D^\mu\Phi - \Phi(D^\mu \Phi)^* \right],
\label{Eq:EoMToyAmu}\\
D^\mu D_\mu\Phi &=& \lambda\Phi,
\label{Eq:EoMToyPhi}\\
|\Phi| &=& 1.
\label{Eq:EoMToylambda}
\end{eqnarray}
Both the action $S_{toy}$ and Eqs.~(\ref{Eq:EoMToyAmu},\ref{Eq:EoMToyPhi},\ref{Eq:EoMToylambda}) are invariant with respect to local gauge transformations
\begin{equation}
\Phi \mapsto e^{-iq\Lambda}\Phi,\qquad
A_\mu\mapsto A_\mu + \partial_\mu\Lambda,
\label{Eq:LocalU1Transf}
\end{equation}
for some arbitrary function $\Lambda$. With an appropriate choice of $\Lambda$ we can always arrange that $\Phi$ is real and positive, in which case the constraint~(\ref{Eq:EoMToylambda}) yields $\Phi = 1$. In this gauge, Eq.~(\ref{Eq:EoMToyPhi}) simplifies to
$$
i q\partial^\mu A_\mu - q^2 A^\mu A_\mu = \lambda.
$$
The real part of this equation fixes the Lagrange multiplier $\lambda = - q^2 A^\mu A_\mu$. More interestingly, the imaginary part of this equation yields the Lorenz gauge condition~\footnote{This gauge condition
is due to Ludvig Lorenz (1829 -- 1891), while the symmetry is due to Hendrik Lorentz (1853 -- 1928). Note the different spelling.}
\begin{equation}
\partial^\mu A_\mu = 0.
\label{Eq:LorentzGauge}
\end{equation}
Using this result in Eq.~(\ref{Eq:EoMToyAmu}) yields the Proca-like equation
\begin{equation}
-\partial_\nu\partial^\nu A^\mu + q^2 A^\mu = 0.
\label{Eq:Proca}
\end{equation}
for the gauge field.

Therefore, instead of enforcing the Lorenz gauge by hand (as is usually done in electromagnetism to obtain a wave equation for the gauge potential), this condition emerges as a consequence of the field equation for the scalar field $\Phi$ and the $U(1)$-gauge adapted to $\Phi$ (such that $\Phi = 1$). Remarkably, this gauge leads naturally to a hyperbolic equation for $A_\mu$. Alternatively, taking into account the Lorenz gauge condition~(\ref{Eq:LorentzGauge}), we can also cast Eq.~(\ref{Eq:Proca}) in first-order form:
\begin{eqnarray}
\partial^\mu A_\mu &=& 0,\\
\partial_\mu A_\nu - \partial_\nu A_\mu &=& F_{\mu\nu},\\
\partial_\nu F^{\mu\nu} &=& -q^2 A^\mu,\\
\partial_{[\alpha} F_{\mu\nu]} &=& 0,
\end{eqnarray}
which yields a symmetric hyperbolic system for $(A_\mu,F_{\mu\nu})$.

This toy model suggests that we should consider a gauge-formulation of GR, in which the role of the scalar field $\Phi$ is replaced by the \ae ther field $u^\alpha$. A gauge-like formulation of GR is provided by the frame formalism.

\section{First-order reformulation of Einstein-{\ae}ther theory in frame variables}
\label{frame_formulation}

Motivated by the example discussed in the previous section, we switch to a frame formulation of Einstein-{\ae}ther theory, in which the spin-2 gravitational field (i.e. the metric) is described by an orthonormal frame $\{ {\bf e}_0, {\bf e}_1, {\bf e}_2, {\bf e}_3 \}$, such that
$$
{\bf g}({\bf e}_\alpha, {\bf e}_\beta) = g_{\mu\nu} e_\alpha{}^\mu e_\beta{}^\nu 
 = \eta_{\alpha\beta},
$$
with $(\eta_{\alpha\beta}) = \mbox{diag}(-1,1,1,1)$. In the following, the Greek indices $\alpha,\beta,\gamma,\delta$ from the beginning of the alphabet denote frame indices, while mid-alphabet letters $\mu,\nu,\ldots$ denote coordinate indices. The frame indices are raised and lowered with the symbol $\eta_{\alpha\beta}$. The coordinate components of the metric can be reconstructed from the frame fields in the following way:
$$
g^{\mu\nu} = \eta^{\alpha\beta} e_\alpha{}^\mu e_\beta{}^\nu.
$$
The Ricci rotation coefficients with respect to the Levi-Civita connection $\nabla$ of ${\bf g}$, defined such that they are antisymmetric in the first two indices, are
$$
\Gamma_{\alpha\beta\gamma} := {\bf g}({\bf e}_\alpha,\nabla_\gamma {\bf e}_\beta)
 = -\Gamma_{\beta\alpha\gamma},
$$
or $\nabla_\gamma {\bf e}_\beta = \Gamma^\alpha{}_{\beta\gamma}{\bf e}_\alpha$.

The transformations that are analogous to~(\ref{Eq:LocalU1Transf}) consist of local Lorentz transformations, 
\begin{equation}
{\bf e}_\alpha\mapsto (\Lambda^{-1})^\beta{}_\alpha {\bf e}_\beta,\qquad
A_\mu \mapsto \Lambda A_\mu\Lambda^{-1} - (\partial_\mu\Lambda)\Lambda^{-1},
\end{equation}
where $\Lambda(x)$ is a Lorentz matrix at each point $x$ of the spacetime manifold (and varying smoothly with $x$), and $A_\mu$ is the matrix-valued connection $1$-form whose components are the matrices $(A_\mu)^\alpha{}_\beta := \Gamma^{\alpha}{}_{\beta\mu}$. Henceforth, we fix part of this freedom by aligning the timelike leg ${\bf e}_0$ of the frame with the \ae ther field ${\bf u}$. As a result of the unit-norm constraint~(\ref{Eq:EoMlambda}), this implies that
$$
{\bf e}_0 = {\bf u}.
$$
In this gauge, one has
$$
J^\alpha{}_\beta = c_1\Gamma_{\beta 0}{}^\alpha 
 + c_2\delta^\alpha{}_\beta\Gamma^\gamma{}_{0\gamma} 
 + c_3\Gamma^\alpha{}_{0\beta} - c_4\delta^\alpha{}_0\Gamma_{\beta00}
$$
and the spatial frame components of Eq.~(\ref{Eq:EoMu}) yield
\begin{equation}
\nabla_\alpha J^\alpha{}_b + c_4 a_\alpha\Gamma^\alpha{}_{0b} = 0,\qquad
b=1,2,3,
\label{Eq:u}
\end{equation}
while the zeroth component fixes the Lagrange multiplier:
\begin{equation}
\lambda = \nabla_\alpha J^\alpha{}_0 + c_4 a^b a_b.
\label{Eq:lambda}
\end{equation}

For the following, it is convenient to use the dyadic formalism of Ref.~\cite{fEhW64}, in which the $24$ independent Ricci rotation coefficients are decomposed into two $3$-vectors
\begin{equation} 
a_b := \Gamma_{b00}, \qquad
\omega_b := -\frac{1}{2}\varepsilon_b{}^{cd}\Gamma_{cd0},\label{acc-w}
\end{equation}
and two $3 \times 3$ matrices
\begin{equation}
K_{ab} := \Gamma_{b0a},\qquad
N_{ab} := \frac{1}{2}\varepsilon_b{}^{cd}\Gamma_{cda},\label{K-N}
\end{equation} 
where from now on $a,b,c,d = 1,2,3$ refer to spatial tetrad indices. $a_b$ is the acceleration of the observers with four-velocity ${\bf e}_0$
(and therefore coincides with the \ae ther acceleration, since ${\bf e}_0={\bf u}$),  and $\omega_b$ is the angular velocity of the spacelike triad relative to a Fermi-propagated frame along such observers. If ${\bf e}_0$ is hypersurface-orthogonal, then $K_{ab} = K_{ba}$ is symmetric and describes the second fundamental form of the hypersurfaces orthogonal to ${\bf e}_0$, while $N_{ab}$ encodes the induced connection of this surface. Below, we will not necessarily assume that ${\bf e}_0 = {\bf u}$ is hypersurface-orthogonal (since that is not generically the case in Einstein-\ae ther theory), so in this case $K_{ab}$ is not necessarily symmetric. Based on this decomposition, one obtains quite naturally a symmetric hyperbolic formulation of Einstein's vacuum equations in General Relativity, see~\cite{lBjB03} and references therein. We will now show that such a formulation can also be obtained in the Einstein-\ae ther theory case.

In terms of the variables $a_b$ and $K_{ab}$ defined above, the $3+1$ split of $J^\alpha{}_\beta$ yields
\begin{eqnarray*}
&& J^0{}_0 = c_2 K,\qquad
J^0{}_b = -c_{14} a_b,\qquad
J^a{}_0 = c_3 a^a,\\
&& J^a{}_b = c_1 K^a{}_b + c_3 K_b{}^a + c_2\delta^a{}_b K,
\end{eqnarray*}
where $K = K^c{}_c$ denotes the trace of $K_{ab}$. 
Up to lower-order terms in the derivatives, Eqs.~(\ref{Eq:u},\ref{Eq:lambda}) respectively yield
\begin{equation}
c_{14} D_0 a_b = c_1 D_a K^a{}_b + c_3 D_a K_b{}^a + c_2 D_b K + l.o.,
\label{Eq:Evola}
\end{equation}
and
\begin{equation}
\lambda = c_2 D_0 K + c_3 D^b a_b + l.o.,
\label{Eq:lambdaPrinc}
\end{equation}
where $D_\alpha = e_\alpha^\mu\partial_\mu$ are the directional derivatives along the tetrad fields. As an instructive example, consider the case where all the $c_i$'s vanish except $c_1$. Eq.~(\ref{Eq:Evola}) then reduces to
$$
D_0 a_b = D_a K^a{}_b + l.o.\,,
$$
which coincides (up to lower-order terms in the derivatives) to what is obtained from the Lorenz gauge condition $D_\gamma\Gamma_{\alpha\beta}{}^\gamma + \Gamma_{\alpha\beta}{}^\gamma\Gamma^\delta{}_{\gamma\delta} = 0$, see~\cite{lBjB03}. Therefore, at least in this example, the \ae ther equation
of motion selects the Lorenz gauge, i.e. it plays the role of the scalar field $\Phi$ in the toy model of Sec. \ref{toy}.

In terms of the variables $K_{ab}$, $N_{ab}$, $a_b$ and $\omega_b$, the modified Einstein equations~(\ref{Eq:EoMg}) can be obtained from the corresponding equations in Refs.~\cite{lBjB03,jBoSlB11}:
\begin{eqnarray}
D_0 K_{ab} - \varepsilon_a{}^{cd} D_c N_{db} &=& (D_a + a_a) a_b
  - \varepsilon_b{}^{cd} N_{ac} a_d + 2\varepsilon_{(a}{}^{cd} K_{b)c} \omega_d 
  +\omega_b \varepsilon_a{}^{cd} K_{cd} + N N_{ab}
 \nonumber \\
&+& \frac{1}{2} \varepsilon_a{}^{df}\varepsilon_b{}^{ce} \left( K_{dc} K_{fe}
    -N_{dc} N_{fe} \right)  - K_a{}^c K_{cb} - N^c{}_a N_{cb} 
 + T^{\text{\ae}}_{ab}
 - \frac{1}{2}\delta_{ab}\left[ (T^{\text{\ae}})^c{}_c + T^{\text{\ae}}_{00} \right],
\label{Eq:KabEvoln}\\
D_0 N_{ab} + \varepsilon_a{}^{cd} D_c K_{db} &=& -(D_a + a_a)\omega_b
  + \varepsilon_b{}^{cd} \left(K_{ac} a_d +N_{ac} \omega_d \right)
  + \varepsilon_a{}^{cd} N_{cb} \omega_d + a_b\varepsilon_a{}^{cd} N_{cd}
 \nonumber \\
&-& N K_{ab} + N^c{}_{a} K_{cb} - N^c{}_b K_{ac}
 +\varepsilon_a{}^{df} \varepsilon_b{}^{ce} N_{dc} K_{fe}
 + \varepsilon_{ab}{}^c T^{\text{\ae}}_{0c},
\label{Eq:NabEvoln}
\end{eqnarray}
where $N$ is the trace of $N_{ab}$. As explained previously, the variables $\omega_b$ appearing on the right-hand sides of these equations are related to the rotational freedom in the choice of the triad fields ${\bf e}_1$, ${\bf e}_2$ and ${\bf e}_3$, and in this work we will assume that they are a priori given functions. Up to lower-order terms, explicit expressions for $T^{\text{\ae}}_{\alpha\beta}$ are obtained from Eq.~(\ref{Eq:TaeDef}),
\begin{eqnarray}
T^{\text{\ae}}_{00} &=& c_{14} D^b a_b + l.o.,\\
T^{\text{\ae}}_{0b} &=& c_{13} D^a K_{(ab)}  + c_2 D_b K + l.o.,\\
T^{\text{\ae}}_{ab} &=& c_{13} D_0 K_{(ab)} + c_2\delta_{ab} D_0 K + l.o.,
\end{eqnarray}
with the notation $K_{(ab)} := (K_{ab} + K_{ba})/2$. Here, we have also used Eq.~(\ref{Eq:u}) in order to simplify the expression for $T^{\text{\ae}}_{0b}$.

Formally, the system~(\ref{Eq:Evola},\ref{Eq:KabEvoln},\ref{Eq:NabEvoln}) would seem a closed evolution system for the variables $(K_{ab},N_{ab},a_b)$; however, one needs to remember that the operators $D_0$ and $D_a$ are directional derivatives along the tetrad fields ${\bf e}_0$ and ${\bf e}_a$, respectively. Therefore, in order to close the system, equations determining the components of the tetrad fields have to be provided. Following~\cite{lBjB03}, we assume
 a given foliation of the spacetime manifold in spacelike hypersurfaces with adapted coordinates $(t,x^i)$ and we decompose the tetrad fields as
\begin{equation}
{\bf e}_0 = \frac{1}{\alpha}\left( \frac{\partial}{\partial t} - \beta^i\frac{\partial}{\partial x^i} \right),\qquad
{\bf e}_b = A_b {\bf e}_0 + B_b^i\frac{\partial}{\partial x^i},
\label{Eq:e0ebDecomposition}
\end{equation}
with $\alpha$ and  $\beta^i$ {denoting the time- and spatial components of the aether field ${\bf u} = {\bf e}_0$,\footnote{Note that in general $\alpha$ and $\beta^i$ are different than the definitions of lapse and shift in the standard $3+1$ decomposition, since ${\bf e}_0$ does not necessarily coincide with the normal vector to the $t = const$ slices.}} and where the fields $A_b$ and $B_b^i$ refer to the components of ${\bf e}_b$ with respect to the non-orthogonal basis of tangent vectors ${\bf e}_0$, $\partial/\partial x^i$. We assume that $\alpha$ and $\beta^i$ are specified by some appropriate gauge conditions, not modifying the principal symbol of the system (for example, they could be a priori specified functions). In contrast to this, the fields $A_b$ and $B_k^i$ are determined by a set of advection equations along the \ae ther field $\bf u$, given by
\begin{eqnarray}
D_0 A_b &=& a_b - B_b^k\frac{\partial}{\partial x^k}\log\alpha 
 - (K_b{}^d + \varepsilon_b{}^{cd}\omega_c) A_d,
\label{Eq:D0Ab}\\
D_0 B_b^k &:=& \frac{1}{\alpha}\left( \frac{\partial}{\partial t} - \pounds_\beta \right) B_b^k
 = -(K_b{}^d + \varepsilon_b{}^{cd}\omega_c) B_d^k,
\label{Eq:D0Bb}
\end{eqnarray}
which follow from the property that the connection is torsion-free~\cite{lBjB03}. The system~(\ref{Eq:D0Ab},\ref{Eq:D0Bb},\ref{Eq:KabEvoln},\ref{Eq:NabEvoln},\ref{Eq:Evola}) provides a closed evolution system for the variables $(A_b,B_b^k,K_{ab},N_{ab},a_b)$, whose hyperbolicity will be analyzed in the next section. Finally, it should be mentioned that this evolution system is subject to several constraint equations. First, there are the Hamiltonian and momentum constraint equations,
\begin{eqnarray}
2D^a n_a &=& -2\omega^a\Omega_a + N^{ab} N_{ab} 
 + \frac{1}{2} K^{ab} K_{ba} - \frac{1}{2} N^{ab} N_{ba}
 - \frac{1}{2} ( K^2 + N^2 ) + T^{\text{\ae}}_{00},
\label{Eq:HamConstraint}\\
D_b K_a{}^b - D_a K &=& -2\varepsilon_a{}^{bc} a_b\Omega_c 
 - \varepsilon_a{}^{bc} K_b{}^d N_{dc} + 2n^b K_{ab} + T^{\text{\ae}}_{0a}
\label{Eq:MomConstraint}
\end{eqnarray}
where we have introduced the shorthand notation $n_a := \varepsilon_a{}^{bc} N_{bc}/2$ and $\Omega_a := \varepsilon_a{}^{bc} K_{bc}/2$. Next, there are analogues constraints on the fields $\Omega_a$ and $N_{ab}$, which read
\begin{eqnarray}
D^a\Omega_a &=& (a^b + 2n^b)\Omega_b,
\label{Eq:OmegaConstraint}\\
D_b N_a{}^b - D_a N &=& 2\varepsilon_a{}^{bc}\omega_b\Omega_c 
 -\varepsilon_a{}^{bc} N_b{}^d N_{dc} - 2\Omega^b K_{ab}.
\label{Eq:NabConstraint}
\end{eqnarray}
Finally, as a further consequence of zero torsion, one obtains the following constraints on the fields $A_b$ and $B_b^i$ (see~\cite{lBjB03}):
\begin{eqnarray}
\varepsilon^{cab} B_a^i \frac{\partial A_b}{\partial x^i} &=& (N^{dc} - \delta^{cd} N) A_d 
 + \varepsilon^{cab} A_a( K_b{}^d + \varepsilon_b{}^{ed}\omega_e - a_b) A_d + 2\Omega^c,
\label{Eq:AbConstraint}\\
\varepsilon^{cab} B_a^i \frac{\partial B_b^k}{\partial x^i} &=& (N^{dc} - \delta^{cd} N) B_d^k 
 + \varepsilon^{cab} A_a(K_b{}^d + \varepsilon_b{}^{ed}\omega_e) B_d^k.
\label{Eq:BbiConstraint}
\end{eqnarray}
Note that unless the \ae ther field ${\bf u}$ is hypersurface orthogonal, such 
that $2\Omega^c = \varepsilon^{abc} K_{ab} = 0$, it is not possible to choose the $t= const$ hypersurface to lie perpendicular to ${\bf u}$; hence, in general one has $A_b\neq 0$. Consequently, 
Eqs.~(\ref{Eq:HamConstraint},\ref{Eq:MomConstraint},\ref{Eq:OmegaConstraint},\ref{Eq:NabConstraint},\ref{Eq:AbConstraint},\ref{Eq:BbiConstraint}) 
should be called ``quasi-constraints'' as in~\cite{lBjB03}, because although they contain only 
directional derivatives along the spatial tetrad legs ${\bf e}_a$, the latter contain partial 
time derivatives, see Eq.~(\ref{Eq:e0ebDecomposition}). However, these equations can be converted 
into genuine constraints without time derivatives of the fields by substituting each appearance of 
$D_b$ by $A_b D_0 + B_b^i\partial_i$, and then using  the evolution equations in order to eliminate all terms involving $D_0$. We will refer to the final equations obtained in this way as the ``constraints''. 

{
Although we do not propose a detailed procedure for solving the initial constraints in this article, we nevertheless make the following observations, which should be relevant for this problem. Another way of turning the ``quasi-constraints''~(\ref{Eq:HamConstraint},\ref{Eq:MomConstraint}) into bona-fide constraints (only depending on initial data for the time evolution) consists of noting that due to the diffeomorphism invariance of the action~\eqref{Eq:AEAction}, there exists a \emph{generalized Bianchi identity}~\cite{Barausse:2011pu,Barausse:2013nwa,Jacobson:2011cc,Seifert:2007fr,Seifert:2006kv,Bergmann:1949zz}
\be
\nabla_\m \left(2 {E^{\m\n}}- u^\mu\AE^\n\right)= \AE_\m\nabla^\n u^\m,
\label{Eq:identity}
\ee
where
\bea
E_{\a\b}&\equiv &  -\frac{1}{2} ( G_{\a\b} - T^{\ae}_{\a\b}),
\\
\AE_\mu& \equiv & \nabla_\a J^{\a}_{\phantom{a}\mu}+c_4 a_\a\nabla_\mu u^\a+\lambda u_\mu  
 = \left(\nabla_\a J^{\a\n}+c_4 a_\a\nabla^\n u^\a\right) \left(g_{\mu\nu}+u_\m u_\n\right).
\eea
(i.e., $E_{\a\b}=0$ and $\AE_\mu=0$ are respectively the Einstein and \ae ther equations).
Expanding the identity~\eqref{Eq:identity} in terms of partial derivatives with respect to a coordinate chart $(t,x^i)$ and the corresponding Christoffel symbols, it is easy to show that the four combinations
\begin{equation}
\label{constraints}
C^\nu\equiv 2 {E^{t\n}}- u^t\AE^\n
\end{equation}
depend on \emph{one less} partial time derivative than the Einstein and \ae ther equations~\cite{Barausse:2011pu}. Since those depend on partial time derivatives of $N_{ab}$, $K_{ab}$ and $a_b$, and they do not contain partial time derivatives of $\alpha,\,\beta^i,\,A_b,\,B^i_b$ and $\omega_b$, it then follows that the combinations $C^\nu$ can only depend on $N_{ab}$, $K_{ab}$ and $a_b$ (but not on their partial time derivatives, nor on $\alpha,\,\beta^i$ and $\omega_b$).\footnote{Note instead that  $C^\nu$ may depend on $A_b,\,B^i_b$, since when applying the argument above, their time derivatives can be eliminated via the advection equations~(\ref{Eq:D0Ab},\ref{Eq:D0Bb}).} Indeed, it can also be verified explicitly that the combinations $T_{\ae}^{t\nu} - u^t\AE^\nu$ only depend on $\nabla_\alpha u_\beta$ and on first-order \emph{spatial} derivatives of $J^\alpha{}_\mu$, which in turn depend algebraically on $a_b$ and $K_{ab}$ (while the Einstein part $G^{t\nu}$ only depends on spatial derivatives of the first and second fundamental forms $(\gamma_{ij},{\cal K}_{ij})$). Furthermore, it is possible to express $K_{ab}$ in terms of $a_b$, ${\cal K}_{ij}$, as well as the spatial components $u_i$ of the \ae ther field and their first-order spatial derivatives.

Consequently, the equations $C^\nu = 0$ yield four constraint equations for the data $(\gamma_{ij},{\cal K}_{ij},u_i,a_i)$, which should be solved on the initial slice $t = 0$ (the tetrad is not required for this, one can solve these constraints in terms of the local spatial coordinates $x^i$). Once these have been solved, one specifies lapse and shift, which allows one to determine completely the components of the $4$-metric and \ae ther field at the initial time $t = 0$. Next, one chooses (e.g. by a Gram-Schmidt method) a tetrad whose timelike leg ${\bf e}_0$ coincides with the \ae ther field, at each point of the initial hypersurface. Finally, one specifies, by a gauge choice, $\omega_b$ (c.f. Eq. (\ref{acc-w})), which determines the time derivative of the triad at the initial slice. $N_{ab}$ is then defined by Eq.~(\ref{K-N}). The constraints given by Eqs.~(\ref{Eq:OmegaConstraint},\ref{Eq:NabConstraint},\ref{Eq:AbConstraint},\ref{Eq:BbiConstraint}) will then be satisfied by construction.

\section{Strong hyperbolicity}
\label{strong_hyper}

In this section we analyze under what conditions on the coupling constants $c_i$ the first-order evolution system~(\ref{Eq:Evola},\ref{Eq:KabEvoln},\ref{Eq:NabEvoln},\ref{Eq:D0Ab},\ref{Eq:D0Bb}) is strongly hyperbolic, such that the associated Cauchy problem is well-posed, at least locally in time. The analysis is performed in several steps, with each reducing the system to one with a smaller number of variables. Roughly speaking, the idea is the following. If we discard all lower-order (undifferentiated) terms in the equations, the evolution equations for the tetrad fields become trivial:
$$
D_0 A_b = 0,\qquad D_0 B_b^k = 0,
$$
while the principal part of the remaining evolution equations~(\ref{Eq:Evola},\ref{Eq:KabEvoln},\ref{Eq:NabEvoln}) gives
\begin{eqnarray}
D_0 K_{ab} &=& +\varepsilon_a{}^{cd} D_c N_{db} + D_a a_b
 + c_{13} D_0 K_{(ab)} - \frac{1}{2} c_{123}\delta_{ab} D_0 K
  - \frac{1}{2} c_{14}\delta_{ab} D^c a_c,
\label{Eq:D0Kab}\\
D_0 N_{ab} &=& -\varepsilon_a{}^{cd} D_c K_{db} - D_a\omega_b 
 + c_{13}\varepsilon_{ab}{}^c D^d K_{(cd)} + c_2\varepsilon_{ab}{}^c D_c K,
\label{Eq:D0Nab}\\
c_{14} D_0 a_b &=& c_1 D^a K_{ab} + c_3 D^a K_{ba} + c_2 D_b K,
\label{Eq:D0ab}
\end{eqnarray}
where we have kept, for the moment, the derivatives of the fields $\omega_b$ in the equations, for future generalizations to more general gauge choices. Note that Eqs.~(\ref{Eq:D0Kab},\ref{Eq:D0Nab},\ref{Eq:D0ab}) are exact in the limit where the fields represent linear perturbations of Minkowski spacetime with a constant \ae ther field; however in general they are only exact up to lower order terms in the derivatives. Next, we note that by applying the derivative operator $D_0$ on both sides of Eq.~(\ref{Eq:D0Kab}), commuting $D_0$ with $D_a$ and using Eqs.~(\ref{Eq:D0Nab},\ref{Eq:D0ab}) in order to eliminate $D_0 N_{ab}$ and $D_0 a_b$ one obtains (up to lower-order terms) a second-order equation for $K_{ab}$, which reads
\begin{eqnarray}
D_0^2 K_{ab} &=& D^c D_c K_{ab} + \left( \frac{c_1}{c_{14}} - 1 \right) D_a D^c K_{cb} 
 + \frac{c_3}{c_{14}} D_a D^c K_{bc} + c_2\left( \frac{1}{c_{14}} - 1 \right) D_a D_b K
 \nonumber\\
 &-& c_{13} D_b D^c K_{(ac)} 
 + c_{13} D_0^2 K_{(ab)}
 + \frac{1}{2}\delta_{ab}\left( c_{13} D^c D^d K_{cd} + c_2 D^c D_c K - c_{123} D_0^2 K \right)
 - \varepsilon_a{}^{cd} D_c D_d\omega_b.
\label{Eq:D02Kab}
\end{eqnarray}
In the following, we will derive sufficient conditions on the principal symbol of this second order system that guarantee that the original, first-order system~(\ref{Eq:Evola},\ref{Eq:KabEvoln},\ref{Eq:NabEvoln},\ref{Eq:D0Ab},\ref{Eq:D0Bb}) is strongly hyperbolic. Besides considerably simplifying the analysis (ending up with a system for the $9$ components of $K_{ab}$ instead of a first-order system with $33$ independent variables), this method will also be useful to compute the characteristic fields of the system and to determine whether they are physical fields (i.e. lying in the kernel of the principal symbol associated with both the constraints and gauge-transformations), constraint-violating fields (i.e. lying in the kernel of the principal symbol associated with gauge transformations but outside the one associated with the constraints) or gauge fields according to the classification of~\cite{gCjPoRoSmT03}.

Before analyzing the hyperbolicity of the system, it is important to remark that Eqs.~(\ref{Eq:Evola},\ref{Eq:KabEvoln},\ref{Eq:NabEvoln}) are not in the standard form with only time derivatives of the fields on one side and only spatial derivatives on the other side. 
The reason is two-fold: on one hand, there are  terms involving the $D_0$ operator on the right-hand side of Eq.~(\ref{Eq:KabEvoln}); on the other hand,
the directional derivative operators $D_a$ contain partial time derivatives, see Eq.~(\ref{Eq:e0ebDecomposition}).
The first issue is easily dealt with by introducing the linear algebraic operator $L$ defined by 
\begin{equation}
L K_{ab} := K_{ab} - c_{13} K_{(ab)} + \frac{1}{2} c_{123}\delta_{ab} K\,,
\label{Eq:LDef}
\end{equation}
which is seen to be invertible as long as $1 - c_{13}\neq 0$ and $2 (1+c_2) + c_{123}\neq 0$\footnote{Note that $L$ can be thought of as a matrix acting on the vector $(K_{ab})$. These conditions then follow from the requirement that the determinant of that matrix be non-zero.}. Assuming that these conditions are satisfied, Eq.~(\ref{Eq:KabEvoln}) can be rewritten in the equivalent form
\begin{equation}
D_0 K_{ab} = L^{-1}\left( \varepsilon_a{}^{cd} D_c N_{db} + D_a a_b
  - \frac{1}{2} c_{14}\delta_{ab} D^c a_c \right)\,,
\label{Eq:D0KabExplicit}
\end{equation}
where no $D_0$-derivatives appear on the right-hand side. To deal with the second issue, in the following we will resort to an elegant, fully covariant definition of hyperbolicity discussed in Refs.~\cite{rG96,oR04}, which does not rely on any particular foliation of spacetime and thus is well-adapted to our formulation in terms of directional derivatives.

\subsection{Reduction to the first-order system for the connection variables $K_{ab}$, $N_{ab}$ and $a_b$}

From now on let us assume that $c_{13}\neq 1$, $c_{14}\neq 0$ and $2(1 + c_2) + c_{123}\neq 0$, such that the operator $L$ defined above is invertible and such that the \ae ther equation~\eqref{Eq:D0ab} is non-degenerate. In this case, the full system~(\ref{Eq:Evola},\ref{Eq:KabEvoln},\ref{Eq:NabEvoln},\ref{Eq:D0Ab},\ref{Eq:D0Bb}) of evolution equations can be written schematically in the form
$$
D_0 U = {\cal A}^b D_b U + {\cal F}(U),
$$
with the $33$-component column vector defined by $U = (A_b,B_b^k,K_{ab},N_{ab},a_b)^T$, where the $33\times 33$ matrices ${\cal A}^1$, ${\cal A}^2$, and ${\cal A}^3$ can be read off the principal part of the equations, and where ${\cal F}(U)$ is a nonlinear function of $U$ that represents the lower-order terms. By defining ${\cal A}^0 := -I$, with $I$ the identity matrix, we can trivially rewrite this system in the form
\begin{equation}
{\cal A}^\alpha D_\alpha U + {\cal F}(U) = 0.
\label{Eq:FirstOrderSystemComplete}
\end{equation}
The principal symbol of this equation is defined by ${\cal A}(k) := {\cal A}^\alpha k_\alpha$, for any co-vector $k_\alpha$. Let us now recall the following definitions from Ref.~\cite{rG96,oR04}, which will allow us to define hyperbolicity even though our system contains directional (as opposed to partial) derivatives:

\begin{definition}
\label{Def:CSH}
A first-order system of the form~(\ref{Eq:FirstOrderSystemComplete}) with an $m$-component state vector $U$ is called {\bf C-strongly hyperbolic} if there exists a co-vector field $n_\alpha$, and, for each $k_\alpha$, an $m\times m$ matrix $h(k)$ depending smoothly on $k_\alpha$, such that
\begin{enumerate}
\item[(i)] $h(k) {\cal A}(k)$ is symmetric for all $k_\alpha$,
\item[(ii)] $h(k) {\cal A}(n)$ is symmetric and positive definite for all $k_\alpha$.
\end{enumerate}
{If the matrix $h(k)$ can be chosen to be independent of $k$, the first-order system~(\ref{Eq:FirstOrderSystemComplete}) is called {\bf C-symmetric hyperbolic}.}
\end{definition}
Since this definition might not be familiar to the reader, we show the relation with the usual definitions of strong and symmetric hyperbolicity for quasi-linear partial differential equations in appendix~\ref{App:Hyperbolicity}. For the applications in this work, it is important to stress that C-symmetric hyperbolicity implies (local in time) well-posedness of the nonlinear Cauchy problem, while a similar result in the strongly hyperbolic case seems to require some additional smoothness conditions (see the discussion at the end of appendix~\ref{App:Hyperbolicity} for details); as far as we are aware of C-strong hyperbolicity as defined above only guarantees the well-posedness of the frozen coefficient problem.

In order to apply definition~\ref{Def:CSH} to our system, we partition the state vector $U$ in the form $U = (E,V)$, with $E = (A_b,B_b^k)$ the components of the tetrad fields and $V = (K_{ab},N_{ab},a_b)$ the connection fields. With respect to this decomposition, the principal symbol has the following block structure:
$$
{\cal A}(k) = -k_0\left( \begin{array}{cc} I & 0 \\ 0 & I \end{array} \right) 
 +  \left( \begin{array}{cc} 0 & 0 \\ 0 & P(\ve{k}) \end{array} \right),
$$
with $P(\ve{k})$ a symbol depending linearly on $\ve{k} := (k_a) = (k_1,k_2,k_3)\in \Real^3$ that can be inferred from the system~(\ref{Eq:D0Kab},\ref{Eq:D0Nab},\ref{Eq:D0ab}). We make the following ansatz for the family of matrices $h(k)$ in definition~\ref{Def:CSH}:
$$
h(k) := \left( \begin{array}{cc} I & 0 \\ 0 & H(\ve{k}) \end{array} \right),
$$
with $H(\ve{k}) = H(\ve{k})^T$ a family of symmetric, positive definite $21\times 21$ matrices to be determined, depending smoothly on $\ve{k}$. Then, taking $n_\alpha := u_\alpha$ to be the co-vector field associated with the \ae ther field (such that ${\cal A}(u) = I$) one finds $h(k) {\cal A}(u) = h(k)$ which, by definition, satisfies condition (ii). Furthermore,
$$
h(k) {\cal A}(k) = -k_0\left( \begin{array}{cc} I & 0 \\ 0 & H(\ve{k}) \end{array} \right) 
 +  \left( \begin{array}{cc} 0 & 0 \\ 0 & H(\ve{k}) P(\ve{k}) \end{array} \right).
$$
which is symmetric provided $H(\ve{k})$ satisfies
\begin{equation}
H(\ve{k}) P(\ve{k}) = P(\ve{k})^T H(\ve{k})
\label{Eq:Symmetrizer}
\end{equation}
for all $\ve{k}\in\Real^3$. (In fact, since $P(\ve{k})$ depends linearly on $\ve{k}$, it is sufficient to assume that $\ve{k}\in S^2$ has unit norm, provided that $H(\ve{k})$ is chosen to depend on the direction of $\ve{k}$, but not on its norm.) Condition~(\ref{Eq:Symmetrizer}) means that the evolution system~(\ref{Eq:D0Kab},\ref{Eq:D0Nab},\ref{Eq:D0ab}) for the connection variables is strongly hyperbolic:

\begin{definition}
\label{Def:SH}
The first-order system~(\ref{Eq:D0Kab},\ref{Eq:D0Nab},\ref{Eq:D0ab}) is called {\bf strongly hyperbolic} if there exists a family of symmetric, positive-definite matrices $H(\ve{k}) = H(\ve{k})^T > 0$ depending smoothly on $\ve{k}\in S^2$ such that Eq.~(\ref{Eq:Symmetrizer}) is satisfied for all $\ve{k}\in S^2$. If the symmetrizer $H = H(\ve{k})$ can be chosen to be independent of $\ve{k}$, then the system is called {\bf symmetric hyperbolic}.
\end{definition}

The condition~(\ref{Eq:Symmetrizer}) implies that $P(\ve{k})$ is diagonalizable (i.e., having a complete set of eigenvectors)  with only real eigenvalues\footnote{This can be seen by noting that Eq.~(\ref{Eq:Symmetrizer}) implies that $P := P(\ve{k})$ is symmetric with respect to the scalar product $(V_1,V_2) := V_1^T H(\ve{k}) V_2$, and thus by the spectral theorem it is diagonalizable with real eigenvalues. Equivalently, Eq.~(\ref{Eq:Symmetrizer}) implies that $H^{1/2}P H^{-1/2}$ (where $H^{1/2}$ is well defined because $H := H(\ve{k})$ is positive definite) is symmetric. If $v$ is an eigenvector for $H^{1/2}P H^{-1/2}$, then $H^{-1/2} v$ is an eigenvector for $P$ with the same eigenvalue. It then trivially follows
that $P$ has a complete set of eigenvectors and is therefore diagonalizable with real eigenvalues.}.  
Conversely, if $P(\ve{k})$ has only real eigenvalues and a complete set of eigenvectors that are arranged in the columns of a matrix $S(\ve{k})$, then $H(\ve{k}) = [S(\ve{k})^{-1} ]^T S(\ve{k})^{-1}$ satisfies Eq.~(\ref{Eq:Symmetrizer}). Therefore, up to the smoothness requirement on $H(\ve{k})$, strong hyperbolicity is equivalent to the principal symbol $P(\ve{k})$ being diagonalizable with only real eigenvalues. 

Summarizing what we have achieved so far, we have shown that the full system of evolution equations~(\ref{Eq:Evola},\ref{Eq:KabEvoln},\ref{Eq:NabEvoln},\ref{Eq:D0Ab},\ref{Eq:D0Bb}) is C-strongly (C-symmetric) hyperbolic if the system~(\ref{Eq:D0Kab},\ref{Eq:D0Nab},\ref{Eq:D0ab}) of the connection variables is strongly (symmetric) hyperbolic. In the next step, we will show that the $21\times 21$ symbol $P(\ve{k})$ can be further reduced to a $9\times 9$ symbol.

\subsection{Reduction to a second-order system for $K_{ab}$}

Further partitioning the state vector $V = (V_1,V_2)$ into the components $V_1 = (K_{ab})$ and $V_2 = (N_{ab},a_b)$, the symbol $P(\ve{k})$ associated with the system~(\ref{Eq:D0Kab},\ref{Eq:D0Nab},\ref{Eq:D0ab}) has the particular block structure
\begin{equation}
P(\ve{k}) = \left( \begin{array}{cc} 0 & Q(\ve{k}) \\ R(\ve{k}) & 0 \end{array} \right),
\label{Eq:PDef}
\end{equation}
with
\begin{eqnarray}
Q(\ve{k}) V_2 &=& L^{-1}\left( \varepsilon_a{}^{cd} k_c N_{db} + k_a a_b 
 - \frac{1}{2} c_{14} \delta_{ab} k^c a_c \right),
 \label{Eq:QSymbol}\\
R(\ve{k}) V_1 &=& \left( \begin{array}{c}
-\varepsilon_a{}^{cd} k_c K_{db} + c_{13} \varepsilon_{abc} k_d K^{(cd)} + c_2\varepsilon_{abc} k^c K
\\
\frac{c_1}{c_{14}} k^a K_{ab} + \frac{c_3}{c_{14}} k^a K_{ba} + \frac{c_2}{c_{14}} k_b K
\end{array} \right),
\label{Eq:RSymbol}
\end{eqnarray}
where we recall that the linear algebraic operator $L$ has been defined in Eq.~(\ref{Eq:LDef}).

The analysis of the symbol $P(\ve{k})$ is greatly simplified by exploiting its special block structure~(\ref{Eq:PDef}), following ideas described in~\cite{ Kreiss2002, oSmT02}. If $\lambda$ is an eigenvalue of $P(\ve{k})$, then there exists $V = (V_1,V_2)\neq (0,0)$ such that
$$
\lambda\left( \begin{array}{c} V_1 \\ V_2 \end{array} \right)
 = \left( \begin{array}{cc} 0 & Q(\ve{k}) \\ R(\ve{k}) & 0 \end{array} \right)
 \left( \begin{array}{c} V_1 \\ V_2 \end{array} \right).
$$
There are two possible type of solutions: either $\lambda = 0$ in which case $V_1$ has to lie in the kernel of $R(\ve{k})$ and $V_2$ in the kernel of $Q(\ve{k})$, or $\lambda \neq 0$, in which case $V_1$ must be an eigenvector of $M(\ve{k}) := Q(\ve{k}) R(\ve{k})$ with eigenvalue $\lambda^2$. Note that $M(\ve{k})$ is precisely the symbol associated with the second-order wave-like equation for $K_{ab}$, Eq.~(\ref{Eq:D02Kab}). Indeed, that equation can be written (up to lower order terms) as $D_0^2 V_1=  M^{ab} D_aD_b V_1$, with $V_1$ the state vector $V_1 = (K_{ab})$, and $M^{ab}$ is the matrix associated to $M(\ve{k})$, i.e. $M(\ve{k}) = M^{ab} k_a k_b$. The following lemma, whose proof is given in the appendix~\ref{App:Proofs}, gives sufficient conditions on the second-order symbol $M(\ve{k})$ for the first-order system to be strongly hyperbolic.

\begin{lemma}
\label{Lem:SecondOrderReduction}
\begin{enumerate}
\item[(a)] Suppose the symbol $M(\ve{k}) = Q(\ve{k}) R(\ve{k})$ is diagonalizable and that all its eigenvalues $0 < \mu_1 < \mu_2 < \ldots < \mu_l$ are strictly positive. Then, $P(\ve{k})$ is diagonalizable with real eigenvalues of the form $0$, $\pm\sqrt{\mu_j}$, $j=1,2,\ldots,l$. (Hence, the system is strongly hyperbolic provided a smooth symmetrizer $H(\ve{k})$ can be constructed.)
\item[(b)] If there exists a family of symmetric, positive-definite matrices $H_1(\ve{k})$ depending smoothly on $\ve{k}\in S^2$ such that
$$
H_1(\ve{k}) M(\ve{k}) = M(\ve{k})^T H_1(\ve{k})
$$
is symmetric and positive definite for all $\ve{k}\in S^2$, then $P(\ve{k})$ is strongly hyperbolic in the sense of definition~\ref{Def:SH}.
\end{enumerate}
\end{lemma}
 
The significance of the lemma relies in the fact that it provides a sufficient condition on the symbol associated with the second-order system for the original system to be strongly hyperbolic. This is clearly a huge simplification, since the former system is $9\times 9$, whereas the latter is $21\times 21$. Explicitly, the symbol $M(\ve{k})$ is given by
\begin{eqnarray}
L M(\ve{k})K_{ab} &=& L K_{ab} + \left( \frac{c_1}{c_{14}} - 1 \right) k_a k^c K_{cb} 
 + \frac{c_3}{c_{14}} k_a k^c K_{bc} + c_2\left( \frac{1}{c_{14}} - 1 \right) k_a k _b K
\nonumber\\
&+& c_{13}\left[ K_{(ab)} - k_b k^c K_{(ac)} - \frac{1}{2}\delta_{ab}(K - k^c k^d K_{cd}) \right].
\label{Eq:MkSymbol}
\end{eqnarray}
A more explicit form for the symbol $M(\ve{k})$ can be obtained by multiplying both sides of the equations to the left with the inverse $L^{-1}$ of $L$; however, it turns out to be simpler to perform this operation after the next step, in which we decompose $K_{ab}$ into scalar, vector and tensor contributions.

\subsection{Decomposition into scalar, vector and tensor blocks}

In order to determine under which conditions on the constants $c_i$'s the hypothesis on the second-order symbol~(\ref{Eq:MkSymbol}) in Lemma~\ref{Lem:SecondOrderReduction} are satisfied, it is convenient to decompose $K_{ab}$ into its components parallel and perpendicular to the unit vector $\ve{k}$. Introducing the  operator $\gamma_a{}^b = \delta_a{}^b - k_a k^b$, which projects on the plane orthogonal to $\ve{k}$, this decomposition reads
\begin{equation}
K_{ab} = k_a k_b K_{kk} + k_a\overline{K}_{k b} + \overline{K}_{a k} k_b + \hat{K}_{ab}
 + \frac{1}{2}\gamma_{ab}(K - K_{kk}),
\label{Eq:KDecomp}
\end{equation}
with the quantities $K_{kk} := k^c k^d K_{cd}$ and $K := \delta^{ab} K_{ab}$ constituting the scalar block, the vectors orthogonal to $\ve{k}$ defined by $\overline{K}_{k b} :=  k^c K_{cd}\gamma^d{}_b$ and $\overline{K}_{a k} := \gamma_a{}^c K_{cd} k^d$ constituting the vector block, and the transverse, trace-less part $\hat{K}_{ab} := \gamma_a{}^c\gamma_b{}^d(K_{cd} - \frac{1}{2}\gamma_{cd}\gamma^{ef} K_{ef} )$. The latter can be further decomposed into symmetric and anti-symmetric parts, the symmetric part $\hat{K}_{(ab)}$ describing the tensor block and the antisymmetric part being dual to the pseudo-scalar $k_a\varepsilon^{abc}\hat{K}_{bc}$. With respect to this decomposition, the eigenvalue problem $\lambda^2 K_{ab} = M(\ve{k}) K_{ab}$ decouples, and one can analyze the conditions of Lemma~\ref{Lem:SecondOrderReduction} separately in each block, which further simplifies the problem. Using  Eqs.~(\ref{Eq:LDef},\ref{Eq:MkSymbol}), the results obtained in each block are the following:
\begin{enumerate}
\item {\bf Tensor and pseudo-scalar blocks}\\
In this case one obtains the set of equations
\begin{equation}
\lambda^2\hat{K}_{(ab)} = \lambda_T^2 \hat{K}_{(ab)},\qquad
\lambda^2\hat{K}_{[ab]} = \hat{K}_{[ab]},
\label{Eq:TensorBlock}
\end{equation}
for the symmetric and anti-symmetric parts of $\hat{K}_{ab}$, respectively, where $\lambda_T^2 := (1 - c_{13})^{-1}$. The system~(\ref{Eq:TensorBlock}) is already in diagonal form and its eigenvalues are equal to   $\lambda_T^2$ and $1$, which are positive and real provided $c_{13} < 1$.

\item {\bf Vector block}\\
In this case one obtains the coupled system
\begin{equation}
\lambda^2\left( \begin{array}{c} K_{k b} \\ K_{b k} \end{array} \right)
 = \left( I + \ve{a}_V \ve{b}_V^T \right)\left( \begin{array}{c} K_{k b} \\ K_{b k} \end{array} \right),
\label{Eq:VectorBlock}
\end{equation}
with $\ve{a}_V$ a column vector and $\ve{b}_V^T$ a row vector given by
$$
\ve{a}_V = \frac{1}{2(1 - c_{13})}\left( \begin{array}{r} 2 - c_{13} \\ c_{13} \end{array} \right),\qquad
\ve{b}_V^T = \left( \frac{c_{13}}{2} + \frac{c_1}{c_{14}} - 1, \frac{c_{13}}{2} + \frac{c_3}{c_{14}}  \right).
$$
In Eq.~(\ref{Eq:VectorBlock}), it is understood that the first component of $\ve{b}_V^T$ acts on the three components of $K_{k b}$ and its second component on the three components of $K_{b k}$, and likewise for $\ve{a}_V$.

\item {\bf Scalar block}\\
In this case one obtains a coupled system, which can be written in a form similar to that of the vector case:
\begin{equation}
\lambda^2\left( \begin{array}{l} K_{kk} \\ K  \end{array} \right)
 = \left( I + \ve{a}_S \ve{b}_S^T \right)\left( \begin{array}{l} K_{kk} \\ K \end{array} \right),
\label{Eq:ScalarBlock}
\end{equation}
with
$$
\ve{a}_S = \frac{1}{(1 - c_{13})}\frac{2}{2 + 2c_2 + c_{123}}
\left( \begin{array}{r} 1 + c_2 \\ 1 - c_{13} \end{array} \right),\qquad
\ve{b}_S^T = \left( \frac{c_{13}}{2} + \frac{c_{13}}{c_{14}} - 1,- \frac{c_{13}}{2} + \frac{c_2}{c_{14}}  - c_2 
\right).
$$
\end{enumerate}

In order to determine under what conditions the vector and scalar blocks are diagonalizable, one can exploit the particular structure of the matrices in each case and use the following simple lemma, whose proof is included in appendix~\ref{App:Proofs} for completeness:

\begin{lemma}
\label{Lem:Auxiliary}
Let $\ve{a},\ve{b}\in \Real^n$ be two non-vanishing constant column vectors in $\Real^n$, and consider the matrix
$$
{\cal M} := I + \ve{a} \ve{b}^T.
$$
Then, ${\cal M}$ is diagonalizable if and only if $\ve{a}$ is not orthogonal to $\ve{b}$. In this case, its eigenvalues are $1 + \ve{a}^T\ve{b}$ (with multiplicity $1$ and eigenvector $\ve{a}$) and $1$ (with multiplicity $n-1$ and eigenvectors orthogonal to $\ve{b}$). Furthermore, the symmetric matrices
$$
{\cal H} := \kappa_0(|\ve{a}|^2 I - \ve{a}\ve{a}^T) + \kappa_1\ve{b}\ve{b}^T,
$$
with positive constants $\kappa_0 > 0$ and $\kappa_1 > 0$, constitute a family of symmetrizers for ${\cal M}$, i.e. they are symmetric, positive-definite and satisfy ${\cal H} {\cal M} = {\cal M}^T {\cal H}$.
\end{lemma}

Applying this lemma to the systems obtained in the tensor, vector and scalar blocks, we conclude that the symbol $M(\ve{k})$ is diagonalizable with strictly positive eigenvalues if and only if
\begin{eqnarray}
\lambda_T^2 &:=& \frac{1}{1 - c_{13}} > 0,
\label{Eq:lambdaT2}\\
\lambda_V^2 &:=& 1 + \ve{b}_V^T\ve{a}_V = \frac{(2 - c_1)c_1 + c_3^2}{2c_{14}(1 - c_{13})} > 0,
\label{Eq:lambdaV2}\\
\lambda_S^2 &:=& 1 + \ve{b}_S^T\ve{a}_S 
 = \frac{c_{123}(2 - c_{14})}{c_{14}(1 - c_{13})(2 + 2c_2 + c_{123})} > 0,
\label{Eq:lambdaS2}
\end{eqnarray}
and $\lambda_V^2 \neq 1 \neq \lambda_S^2$. If these conditions are satisfied, one can use the previous Lemma again to construct a symmetrizer $H_1(\ve{k})$ for the second-order symbol $M(\ve{k})$, defined by
\begin{eqnarray*}
(K,K)_1 &:=& K^{ab} H_1(\ve{k}) K_{ab} \\
  &=& \hat{K}^{ab}\hat{K}_{ab} 
 + \left( \begin{array}{c} K^{k b} \\ K^{b k} \end{array} \right)^T
 \left( |\ve{a}_V|^2 I - \ve{a}_V\ve{a}_V^T + \ve{b}_V\ve{b}_V^T \right)
 \left( \begin{array}{c} K_{k b} \\ K_{b k} \end{array} \right)
 + \left( \begin{array}{c} K_{kk} \\ K \end{array} \right)^T
 \left(  |\ve{a}_S|^2 I - \ve{a}_S\ve{a}_S^T + \ve{b}_S\ve{b}_S^T \right)
 \left( \begin{array}{c} K_{kk} \\ K \end{array} \right).
\end{eqnarray*}
By construction, the matrix $H_1(\ve{k})$ defined in this way is symmetric, positive-definite and satisfies the condition (b) of Lemma~\ref{Lem:SecondOrderReduction}. Furthermore, $H_1(\ve{k})$ depends smoothly on the vector $\ve{k}$, since the projections of $K_{ab}$ onto its pieces parallel and orthogonal to $\ve{k}$ are smooth (in particular, note that the projection operator $\gamma_a{}^b = \delta_a{}^b - k_a k^b$ depends smoothly on $\ve{k}$) and since the components of the vectors $\ve{a}_V,\ve{b}_V,\ve{a}_S,\ve{b}_S$ only depend on the constants $c_i$. Therefore, it follows from Lemma~\ref{Lem:SecondOrderReduction} that the first-order system with symbol $P(\ve{k})$ is strongly hyperbolic, provided the conditions~(\ref{Eq:lambdaT2},\ref{Eq:lambdaV2},\ref{Eq:lambdaS2}) and $\lambda_V^2 \neq 1\neq \lambda_S^2$ are fulfilled. This is the main result of this paper, whose implications will be further discussed in the conclusion section. Before doing so, however, it is instructive to determine the eigenvalue-eigenvector pairs associated with the symbol $P(\ve{k})$ since (at least in the regime of small amplitude, high-frequency perturbations) they describe the propagation speeds and modes of the system.

\subsection{Characteristic speed and fields and their physical interpretation}

Before proceeding, we briefly note that the quantities $\lambda_T^2$, $\lambda_V^2$ and $\lambda_S^2$ defined in~(\ref{Eq:lambdaT2},\ref{Eq:lambdaV2},\ref{Eq:lambdaS2}) coincide precisely with the squared propagation speed $s_2^2$, $s_1^2$ and $s_0^2$, respectively, given in  Eqs.~(\ref{s2},\ref{s1},\ref{s0}). 
This is expected, because, as already discussed, the speeds $s_i$ ($i=0,1,2$) regulate the propagation of the \emph{physical} spin-$i$ modes of the theory, on flat backgrounds. 

From Lemma~\ref{Lem:Auxiliary}, the eigenfields of the second-order symbol $M(\ve{k})$ 
corresponding to the eigenvalues $\lambda_T^2$, $\lambda_V^2$ and $\lambda_S^2$
are given by
\begin{eqnarray}
\hbox{$\lambda_T^2 = s_2^2$ (two modes)} &:& K^{(T,1)}_{ab} = e_a e_b - f_a f_b, \quad
 K^{(T,2)}_{ab} = 2e_{(a} f_{b)},
\label{Eq:KT12}\\
\hbox{$\lambda_V^2 = s_1^2$ (two modes)} &:& K^{(V,1)}_{ab} = (2 - c_{13}) k_a e_b + c_{13} e_a k_b,\quad
K^{(V,2)}_{ab} = (2 - c_{13}) k_a f_b + c_{13} f_a k_b,
\label{Eq:KV12}\\
\hbox{$\lambda_S^2 = s_0^2$ (one mode)} &:& K^{(S)}_{ab}  = (1 + c_2) k_a k_b - \frac{1}{2} c_{123}\gamma_{ab},
\label{Eq:KS}
\end{eqnarray}
with $e_b$, $f_b$ two mutually orthogonal unit vectors such that $\{ \ve{k}, \ve{e}, \ve{f} \}$ forms an oriented orthonormal basis of $\Real^3$, i.e. $\varepsilon_{abc} = 6k_{[a} e_b f_{c]}$. The corresponding eigenfields of the first-order symbol $P(\ve{k})$ can be constructed using the method described in the proof (given in Appendix \ref{App:Proofs}) of Lemma~\ref{Lem:SecondOrderReduction}(a), which yields the eigenvalue-eigenvector pairs
\begin{eqnarray}
\hbox{$\pm\lambda_T$ (four modes)} &:& V_\pm^{(T,1)} = (K^{(T,1)}_{ab}, 
\pm\lambda_T^{-1} R(\ve{k}) K^{(T,1)}_{ab})
 = (K^{(T,1)}_{ab},\mp\lambda_T^{-1}K^{(T,2)}_{ab},0),\nonumber\\
 && V_\pm^{(T,2)} = (K^{(T,2)}_{ab}, \pm\lambda_T^{-1} R(\ve{k}) K^{(T,2)}_{ab})
 = (K^{(T,2)}_{ab},\pm\lambda_T^{-1}K^{(T,1)}_{ab},0),
\label{Eq:vT12}\\
\hbox{$\pm\lambda_V$ (four modes)} &:& V_\pm^{(V,1)} = (K^{(V,1)}_{ab}, 
\pm\lambda_V^{-1} R(\ve{k}) K^{(V,1)}_{ab})
 = \left( K^{(V,1)}_{ab}, \mp\frac{c_{13}}{\lambda_V} k_a f_b, 
 \pm\frac{(2-c_1)c_1 + c_3^2}{\lambda_V c_{14}} e_b \right),\nonumber\\
 && V_\pm^{(V,2)} = (K^{(V,2)}_{ab}, \pm\lambda_V^{-1} R(\ve{k}) K^{(V,2)}_{ab})
 = \left( K^{(V,2)}_{ab}, \pm\frac{c_{13}}{\lambda_V} k_a e_b, 
  \pm\frac{(2-c_1)c_1 + c_3^2}{\lambda_V c_{14}} f_b \right),
\label{Eq:vV12}\\
\hbox{$\pm\lambda_S$ (two mode)} &:& 
 V_\pm^{(S)} = (K^{(S)}_{ab}, \pm\lambda_S^{-1} R(\ve{k}) K^{(S)}_{ab})
 = (K^{(S)}_{ab}, \pm \lambda_S^{-1} e_{[a} f_{b]},\pm\lambda_S^{-1} c_{14}^{-1} k_b ),
\label{Eq:vS}
\end{eqnarray}
where we recall the notation $V = (K_{ab},N_{ab},a_b)$ and that the symbol $R(\ve{k})$ is defined in Eq.~(\ref{Eq:RSymbol}).

The remaining eigenfields of the second-order symbol $M(\ve{k})$ have eigenvalues $1$. According to Lemma~\ref{Lem:Auxiliary}, they are explicitly given by $2e_{[a} f_{b]}$; by two non-trivial linear combinations of $K_{kb}$ and $K_{bk}$ orthogonal to the vector $\ve{b}_V$; and by a non-trivial linear combination of $K_{kk}$ and $K$ orthogonal to $\ve{b}_S$. As we show now, these fields correspond to constraint-violating modes. In order to do so, we consider the symbol associated with the momentum constraint~(\ref{Eq:MomConstraint}):
$$
k^b \left[ K_{ab} - c_{13} K_{(ab)} - (1 + c_2)\delta_{ab} K \right] = 0.
$$
In the scalar and vector blocks, respectively, this yields
$$
(1 - c_{13}) K_{kk} - (1 + c_2)K = 0,\qquad
c_{13} K_{kb} - (2 - c_{13}) K_{bk} = 0,
$$
while there are no restrictions in the tensor and pseudo-scalar blocks. We see that these equations are precisely satisfied for the eigenvectors proportional to $\ve{a}_S$ and $\ve{a}_V$, respectively. Therefore, the five eigenvectors defined in Eqs.~(\ref{Eq:KT12},\ref{Eq:KV12},\ref{Eq:KS}), representing the physical modes, lie in the kernel of the symbol associated with the momentum constraint, as expected. In contrast to this, the vector and scalar modes propagating with speed $1$ are orthogonal to $\ve{b}_V$ and $\ve{b}_S$, respectively, and hence they cannot be parallel to $\ve{a}_S$ or $\ve{a}_V$. Consequently, the vector and scalar modes propagating with speed $1$ are \emph{constraint-violating modes}.

Next, we analyze the eigenfield $2e_{[a} f_{b]}$ in the pseudo-scalar block, which also propagates with speed $1$ but lies in the kernel of the symbol associated with the momentum constraint. The corresponding eigenvectors of the first-order symbol $P(\ve{k})$ are
\begin{eqnarray*}
\left( 2e_{[a} f_{b]}, \pm R(\ve{k}) 2e_{[a} f_{b]} \right) 
 = \left( 2e_{[a} f_{b]}, \mp (e_a e_b + f_a f_b), 0 \right),
\end{eqnarray*}
which do not lie in the kernel associated with the constraint equation~(\ref{Eq:NabConstraint}):
$$
N_{bk} = 0,\qquad N_{kk} - N = 0,
$$
whereas the physical modes defined in Eqs.~(\ref{Eq:vT12},\ref{Eq:vV12},\ref{Eq:vS}) do. Therefore, all the modes (with the exception of the modes $V_\pm^{(T,1)}$ and $V_\pm^{(T,2)}$ when $c_{13}=0$) propagating with speed $1$ are constraint-violating. For completeness, one may also consider the symbol associated with the Hamiltonian constraint~(\ref{Eq:HamConstraint}), which is
$$
\varepsilon^{abc} k_a N_{bc} - c_{14} k^b a_b = 2e^{[a} f^{b]} N_{ab} - c_{14} a_k = 0,
$$
and is automatically satisfied by the physical modes defined in Eqs.~(\ref{Eq:vT12},\ref{Eq:vV12},\ref{Eq:vS}).
 
The remaining $15$ eigenvectors of the first-order system propagate with zero speed. They correspond to the $12$ tetrad fields $A_b$ and $B_b^k$ and to the $3$ independent vectors lying in the kernel of the symbol $Q(\ve{k})$, as discussed in Lemma~\ref{Lem:SecondOrderReduction}(a). The 
kernel of $Q(\ve{k})$ is easily shown to be of the form
$$
V = (K_{ab},N_{ab},a_b) = (0,k_a w_b,0),
$$
with $w_b$ an arbitrary vector, and they correspond to gauge-modes because they can be eliminated by an appropriate choice of the angular velocity $\omega_a$, see Eq.~(\ref{Eq:D0Nab}). This concludes our discussion of the characteristic speeds and fields of the system.

\section{Symmetric hyperbolic formulations}
\label{sym_hyper}

In this section, we show that by taking suitable combinations of the evolution and (quasi-)constraint equations, it is possible to recast the evolution equations into symmetric (instead of merely strongly) hyperbolic form. To simplify the analysis, in the following we focus on the three-parameter space
\begin{equation}
c_1 = \Delta s_1^2,\qquad
c_2 = \Delta \Gamma,\qquad
c_{13} = 0,\qquad
c_{14} = \Delta,
\label{Eq:cChoice}
\end{equation}
with $\Delta$, $s_1$ and $\Gamma$ real parameters satisfying $s_1\neq 0$ and $\Delta\Gamma > -2/3$ (so that the evolution equations and symmetrizer below are well-defined). In terms of the new parametrization, the squared propagation speeds~(\ref{Eq:lambdaT2},\ref{Eq:lambdaV2},\ref{Eq:lambdaS2}) are given by
\begin{equation}
\lambda_T^2 = 1,\qquad
\lambda_V^2 = s_1^2,\qquad
\lambda_S^2 = \frac{(2 - \Delta)\Gamma}{2 + 3\Delta\Gamma},
\label{Eq:lambdas}
\end{equation}
and the post-Newtonian parameters~(\ref{eq:alpha1},\ref{eq:alpha2}) reduce to
\begin{equation}
\alpha_1 = -4\Delta,\qquad
\alpha_2 = \Delta\frac{1 - (1 - 2\Delta)\Gamma}{(2-\Delta)\Gamma}.
\end{equation}
Hence, the parameter choices~(\ref{Eq:cChoice}) are compatible with the observational constraints obtained  in section~\ref{Sec:AE} by saturating the bound on $\alpha_1$, provided that $|\Delta| \lesssim 0.25\times 10^{-4}$ and $|\Gamma- 1|\lesssim  8\times 10^{-3}$ (and $s_1 \geq 1$ and $\Gamma-1 \geq 2 \Delta/(1 - 2 \Delta)$ to ensure $s_i^2 \geq 1$ and thus satisfy the Cherenkov bound). Note that the bounds from gravitational wave generation, though yet to be worked out in detail in this region of the parameter space, should be satisfied. Indeed, if $s_1\sim 1$, then $|c_1-c_3|\sim {\cal O}(10^{-4})$; while if $s_1\gg 1$ and $\Delta s_1^2\gg1$, then $|c_1-c_3|\gg 1$, and the theory should reproduce the GR predictions for gravitational wave generation in both limits as discussed in Sec.~\ref{Sec:AE}. Note that our parametrization can also cover the two-dimensional parameter space $(c_2,c_1-c_3)$ obtained in section~\ref{Sec:AE} by setting  $0\neq|c_3-c_4|\lesssim 10^{-7}$. Here, that corresponds to taking $0\neq |\Delta|\lesssim 10^{-7}$.

To obtain a symmetric hyperbolic system, we use the momentum quasi-constraint equation~(\ref{Eq:MomConstraint}) to eliminate the divergence term $D^a K_{ba}$ in the evolution equation for $a_b$. Furthermore, for the sake of gaining flexibility to achieve a symmetric symbol, we use the Hamiltonian quasi-constraint equation~(\ref{Eq:HamConstraint}) to modify the right-hand side of the evolution equation for $K_{ab}$. After these operations, the principal part of the evolution equations~(\ref{Eq:D0Kab},\ref{Eq:D0Nab},\ref{Eq:D0ab}) becomes
\begin{eqnarray}
D_0 K_{ab} &=& +\varepsilon_a{}^{cd} D_c N_{db} + D_a a_b
 -\frac{\Delta}{2 + 3\Delta\Gamma}\delta_{ab} 
 \left[ (\Gamma - \sigma)\varepsilon^{fcd} D_f N_{cd} + (1 + \Gamma + \Delta\sigma) D^c a_c \right],
\label{Eq:D0KabBis}\\
D_0 N_{ab} &=& -\varepsilon_a{}^{cd} D_c K_{db} + \Delta\Gamma\varepsilon_{ab}{}^c D_c K,
\label{Eq:D0NabBis}\\
D_0 a_b &=& s_1^2 D^a K_{ab} +\left[ \Gamma - s_1^2(1 + \Delta\Gamma) \right] D_b K,
\label{Eq:D0abBis}
\end{eqnarray}
with the new real parameter $\sigma$ being the coefficient determining the considered linear combination  between the evolution equation for $K_{ab}$ and the Hamiltonian quasi-constraint. For the particular choice $\Delta = 0$ and $\Gamma = s_1^2 > 0$, the system~(\ref{Eq:D0KabBis},\ref{Eq:D0NabBis},\ref{Eq:D0abBis}) simplifies considerably, and it is a simple task to verify that it is symmetric hyperbolic with respect to the ($\ve{k}$-independent) symmetrizer $H$ defined by
\begin{equation}
V^T H V := K^{ab} K_{ab} + N^{ab} N_{ab} + \frac{1}{s_1^2} a^b a_b,\qquad
V = (K_{ab},N_{ab},a_b)^T,
\label{Eq:SymHypSymmetrizer}
\end{equation}
that is, its principal symbol $P(\ve{k})$ satisfies $V_1^T H P(\ve{k}) V_2 = V_2^T H P(\ve{k}) V_1$ for all state vectors $V_1$ and $V_2$ and all $\ve{k}\in S^2$. More generally, one can show that Eq.~(\ref{Eq:SymHypSymmetrizer}) provides a symmetrizer for the system~(\ref{Eq:D0KabBis},\ref{Eq:D0NabBis},\ref{Eq:D0abBis}), provided that the following relations hold:
$$
\Gamma = -\frac{\Gamma - \sigma}{2 + 3\Delta\Gamma},\qquad
\frac{\Gamma}{s_1^2} - (1 + \Delta\Gamma) = -\Delta\frac{1 + \Gamma + \Delta\sigma}{2 + 3\Delta\Gamma}.
$$
The first one can always be satisfied by defining
\begin{equation}
\sigma := 3\Gamma(1 + \Delta\Gamma),
\label{Eq:sigmaDef}
\end{equation}
while the second reduces the dimensionality of the parameter space $(s_1,\Delta,\Gamma)$ from three to two; e.g. the second relation is satisfied if we define $s_1^2$ by
\begin{equation}
s_1^2 := \Gamma\frac{2 + 3\Delta\Gamma}{1 + \Delta\Gamma 
 + (1 - \Delta)(1 + 3\Delta\Gamma + 3\Delta^2\Gamma^2)}.
\label{Eq:s1Def}
\end{equation}
We conclude that the system~(\ref{Eq:D0KabBis},\ref{Eq:D0NabBis},\ref{Eq:D0abBis}) is symmetric hyperbolic for any choice for $\Delta$ and $\Gamma$ such that the right-hand side of Eq.~(\ref{Eq:s1Def}) is strictly positive, provided that $\sigma$ and $s_1^2$ are defined by Eqs.~(\ref{Eq:sigmaDef},\ref{Eq:s1Def}), respectively. This shows that there is a (at least) two-dimensional parameter space of Einstein-\ae ther theory whose evolution equations can be cast into symmetric hyperbolic form. This parameter space contains the values $(\Delta,\Gamma)$ for which $0\leq \Delta\leq 1$ and $\Gamma > 0$, and hence it also contains the open region in the $(\Delta,\Gamma)$-plane defined by $|\Delta| \lesssim 0.25\times 10^{-4}$ and $|\Gamma- 1|\lesssim  8\times 10^{-3}$, which is compatible with the solar system constraints when saturating the $\alpha_1$ bound. Note indeed that for small $|\Delta|$ and $|\Gamma-1|$, Eq.~(\ref{Eq:s1Def}) gives $s_1^2=1+(\Gamma-1)+{\cal O}[(\Gamma-1)^2,\Delta^2]$, which is enough to satisfy the experimental bounds discussed above (from Cherenkov radiation and gravitational wave emission) if $\Gamma>1$. Note however that the condition Eq.~(\ref{Eq:s1Def}) does not allow us to prove symmetric hyperbolicity in the case in which $s_1$ diverges (which is a necessary condition to cover the limit $|c_1-c_3|\to\infty$, also allowed by the experimental bounds when one saturates the $\alpha_1$ constraint).

By adding a term proportional to $K^2$ to the right-hand side of Eq.~(\ref{Eq:SymHypSymmetrizer}), one can obtain an even larger class of symmetric hyperbolic systems for an open set of parameters $(s_1,\Delta,\Gamma)$ in $\Real^3$, characterized by the requirement that
\begin{equation}
\frac{(1 + 3s_0^2)^2}{2s_1^2}\Gamma^2 + (1 - 3s_0^2)\Gamma^2 - 4s_0^2\Gamma 
 - \frac{s_0^2}{3}(1 - 3s_0^2) > 0,\qquad
s_0^2 = \frac{(2 - \Delta)\Gamma}{2 + 3\Delta\Gamma}.
\end{equation}
When $(\Delta,\Gamma)\approx (0,1)$ one has $s_0^2\approx 1$, and this set restricts the value of $s_1^2$ to be approximately smaller than $3/2$, so one can still not access the regime where $\Delta s_1^2$ is very large, which is in principle also compatible with the experimental bounds, as discussed above.

As for the two-dimensional parameter space $(c_2,c_1-c_3)$ obtained in section~\ref{Sec:AE} by setting  $0\neq|c_3-c_4|\lesssim 10^{-7}$, that is fully included in our analysis since $|\Delta|$ can be made arbitrarily small while still satisfying Eq.~(\ref{Eq:s1Def}). Note also that unless $c_2$ and $c_3$ are also ${\cal O}(\Delta)$ (which can be avoided simply by choosing $\Gamma \sim c_2/\Delta\sim {\cal O}(0.1)/\Delta$), the spin-0 and spin-1 squared propagation speeds automatically become $\sim \Gamma \sim {\cal O}(0.1)/\Delta \sim 10^6$. As already mentioned in section~\ref{Sec:AE}, this renders the scalar and vector polarizations close to non-dynamical, and therefore more likely to pass binary pulsar bounds.

\section{Conclusions}
\label{conclusions}
We have analyzed the well-posedness of the Cauchy (initial value) problem in Einstein-\ae ther theory. In the standard metric formulation of the theory it is far from clear that the evolution equations can be cast in hyperbolic form, since they are second order in both the metric fields and the \ae ther vector field. This is worrisome as well-posedness is a fundamental ingredient for the predictive power of the theory, meaning that for a given set of initial data there exists (at least locally in time) a unique solution that depends continuously on the data. In particular, well-posedness is a necessary requirement to ensure stability and convergence of numerical initial value evolutions. Although well-posedness is relatively easy to prove for linear perturbations on flat spacetime, since the linear equations reduce to a system of decoupled wave equations~\cite{Jacobson:2004ts} for the spin-0, spin-1 and spin-2 modes of the theory, a generalization to the full system of field equations was so far lacking. We have succeeded in showing that a first-order reformulation of Einstein-\ae ther theory in terms of projections onto a tetrad frame quite naturally leads to strongly hyperbolic evolution equations, as long as the (squared) propagation speeds $s_0^2$, $s_1^2$ and $s_2^2$ are strictly positive and finite with $s_0^2 \neq 1 \neq s_1^2$. The covariant notion of strong hyperbolicity employed in this article ensures (modulo technical smoothness requirements, see Appendix~\ref{App:Hyperbolicity}) the local in time well-posedness of the vacuum initial value (Cauchy) problem in Einstein-\ae ther theory in this region of the coupling constants' parameter space. Furthermore, by suitably modifying the evolution equations by means of the constraints, we have obtained a three-parameter family of formulations for which $s_2^2 = 1$ and $s_0^2$ and $s_1^2$ lie close enough to $1$, which is symmetric (rather than merely strongly) hyperbolic, and in this case (local in time) well-posedness of the Cauchy problem follows without any additional assumptions.

At least some of the conditions on the propagation speeds are easy to understand on physical grounds. Strict positiveness of $s_0^2$, $s_1^2$ and $s_2^2$ is needed to ensure absence of ghosts and gradient instabilities (c.f. Refs.~\cite{Jacobson:2004ts,Garfinkle:2011iw} and discussion in section~\ref{Sec:AE}). At the level of the evolution equations, a negative value of $s_0^2$, $s_1^2$ or $s_2^2$ would imply catastrophic, unbounded frequency-dependent instabilities of the solutions. Similarly, if any of the propagation speeds diverges, the field equations cannot be strongly hyperbolic (not even in flat space), as the degree of freedom whose speed diverges becomes non-dynamical and satisfies an elliptic equation in flat space~\cite{Jacobson:2004ts}. Therefore, the system can at best be elliptic-hyperbolic. It is instead more difficult to intuitively make sense of the requirement that the spin-0 and spin-1 speeds should be different than one. If equal to one, it follows from our analysis in section~\ref{strong_hyper} that the evolution equations are only weakly hyperbolic, a property that (at the nonlinear level) usually leads to frequency-dependent instabilities~\cite{KL89,Sarbach:2012pr}. Nevertheless, it seems very likely that these special cases still yield a well-posed Cauchy problem if the constraints are taken into account (indeed the restrictions $s_0^2 \neq 1 \neq s_1^2$ go away in our symmetric hyperbolic subfamilies for which the constraints have been used). This could be analyzed in a systematic and elegant way based on the recent method introduced in Ref.~\cite{fAoR18}. We will leave this problem aside for future work. Nevertheless, the experimental constraints on the coupling constants (discussed in section~\ref{Sec:AE}) show that while the spin-2 speed needs to be very close to unity (to within ${\cal O}(10^{-15})$), the spin-0 and spin-1 speeds are typically different from one. Therefore, our proof of the well-posedness of the Cauchy problem applies to the whole viable parameter space of the coupling constants.

\begin{acknowledgments}
We would like to warmly thank Luis Lehner for providing insightful comments about this work.
We also thank Ted Jacobson, Thomas Sotiriou and Diego Blas for useful discussions 
about Lorentz violating gravity, and Oscar Reula for fruitful discussions about covariant definitions of strong hyperbolicity and their implications for the well-posedness of the Cauchy problem. This project has received funding from  the European Research Council (ERC) under the European Union's Horizon 2020 research and innovation programme (grant agreement No.~GRAMS-815673; project title ``GRavity from Astrophysical to Microscopic Scales''); 
from the European Union's Horizon 2020 research and innovation programme under the
Marie Sklodowska-Curie grant agreement No.~690904; from the CONACyT Network Project No.~294625 ``Agujeros Negros y Ondas Gravitatorias"; and from a CIC grant to Universidad Michoacana de San Nicol\'as de Hidalgo.
This research was supported in part by Perimeter Institute for Theoretical Physics. Research at Perimeter Institute is supported by the Government of Canada through the Department of Innovation, Science and Economic Development Canada and by the Province of Ontario through the Ministry of Economic Development, Job Creation and Trade.
\end{acknowledgments}

\appendix
\section{Equivalent definitions of strong hyperbolicity}
\label{App:Hyperbolicity}

Consider a first-order quasi-linear system of the form of Eq.~(\ref{Eq:FirstOrderSystemComplete}),
\begin{equation}
{\cal A}^\alpha D_\alpha U + {\cal F}(U) = 0,
\label{Eq:SHSystem}
\end{equation}
with corresponding principal symbol ${\cal A}(k) := {\cal A}^\alpha k_\alpha$. Suppose that this system is C-strongly hyperbolic in the sense of definition~\ref{Def:CSH}. In this appendix we reproduce parts of the arguments in~\cite{oR04}, which show that this definition is equivalent to the usual definition of strong hyperbolicity. To this purpose, let $T^\alpha$ denote any vector field such that $n_\alpha T^\alpha = 1$, and let $h_\alpha{}^\beta := \delta_\alpha{}^\beta - n_\alpha T^\beta$ denote the projection operator along $n_\beta$ onto the subspace of co-vectors that annihilate $T^\alpha$ (i.e. $h_\alpha{}^\beta n_\beta = 0$ and $T^\alpha h_\alpha{}^\beta = 0$). Inserting $\delta_\alpha{}^\beta = n_\alpha T^\beta + h_\alpha{}^\beta$ into Eq.~(\ref{Eq:SHSystem}), this equation can be rewritten as
\begin{equation}
T^\beta D_\beta U = -{\cal A}(n)^{-1}\left[ {\cal A}^\alpha h_\alpha{}^\beta D_\beta U + {\cal F}(U) \right]
 =: \tilde{\cal A}^\beta D_\beta U + \tilde{\cal F}(U),
\label{Eq:SHSystemBis}
\end{equation}
where we have used the fact that, because of the second condition of definition~\ref{Def:CSH}, the matrix ${\cal A}(n)$ is invertible. The symbol of the operator on the right-hand side of Eq.~(\ref{Eq:SHSystemBis}) is defined by
\begin{equation}
\tilde{\cal A}(k) := \tilde{\cal A}^\alpha k_\alpha = -{\cal A}(n)^{-1} {\cal A}^\alpha k_\alpha,
\label{Eq:SHSymbol}
\end{equation}
for all co-vectors $k_\alpha$ satisfying $k_\alpha T^\alpha = 0$, and as a consequence of definition~\ref{Def:CSH}, it possesses the smooth symmetrizer
$$
\tilde{\cal H}(k) := h(k) {\cal A}(n) = \tilde{\cal H}(k)^T > 0,\qquad k_\alpha T^\alpha = 0.
$$

To establish the relation with the usual definition of strong hyperbolicity, suppose first that $n_\alpha$ is hypersurface-orthogonal, such that (at least locally) $n_\alpha = -N D_\alpha t$, for some functions $N$ and $t$. Introduce local coordinates $x^1,x^2,x^3$ on the $t = const$ hypersurfaces which are transported along the vector field $T^\alpha$, such that $T^\beta D_\beta = \partial/\partial t$. Since $h_\alpha{}^\beta D_\beta t = -N^{-1} h_\alpha{}^\beta n_\beta = 0$, the differential operator $\tilde{\cal A}^\beta D_\beta$ is tangent to the $t = const$ hypersurfaces, and Eq.~(\ref{Eq:SHSystemBis}) reduces to a strongly hyperbolic quasi-linear partial differential equation with associated principal symbol~(\ref{Eq:SHSymbol}). In this case, local in time well-posedness of the Cauchy problems follows from standard theorems, see for instance~\cite{Taylor99c}.

If $n_\alpha$ is not hypersurface-orthogonal we take a point $p\in M$ on the manifold and approximate $n_\alpha$ by a different co-vector field $\tilde{n}_\alpha$ which is hypersurface-orthogonal in an open neighbourhood ${\cal U}\subset M$ of $p$ such that $\left. \tilde{n}_\alpha \right|_p = \left. n_\alpha \right|_p$. According to Proposition~1 in~\cite{oR04}, the system~(\ref{Eq:SHSystem}) is also C-hyperbolic with respect to $\tilde{n}_\alpha$ provided ${\cal U}$ is chosen small enough, and hence the arguments above show that the system~(\ref{Eq:SHSystemBis}) with $n_\alpha$ replaced by $\tilde{n}_\alpha$ is again strongly hyperbolic.

A relevant question is whether or not the new symmetrizer $\tilde{\cal H}(k)$ (with $n_\alpha$ replaced by $\tilde{n}_\alpha$) is still smooth in $k$, which is required to apply the standard theorems in the variable-coefficient or the quasi-linear cases. Unfortunately, we do not know of any general results which guarantee this property, and hence in the case in which the co-vector field $n_\alpha$ is not hypersurface-orthogonal one can \emph{a priori} only guarantee well-posedness of the frozen coefficient problems. However, there is an important special case in which local in time well-posedness for the quasi-linear problem \emph{does follow}, namely when the system is symmetric hyperbolic, in which case the symmetrizers $h(k)$ and ${\cal H}(k)$ are independent of $k$.

\section{Proofs of Lemma~\ref{Lem:SecondOrderReduction} and~\ref{Lem:Auxiliary}}
\label{App:Proofs}

In this appendix we provide the proofs for the two technical lemmas used in Section~\ref{strong_hyper}.\\

\proofof{Lemma~\ref{Lem:SecondOrderReduction}} For convenience, we rewrite the vector $(V_1,V_2)$ as $(v,w)$ and denote by $V$ and $W$ the vector spaces $v$ and $w$ live in, such that $\dim V =: n \leq m := \dim W$ and $R(k): V\to W$, $Q(k): W\to V$. In order to prove (a), we first note that the hypothesis implies that $Q(k)$ and $R(k)$ have full rank, since otherwise $M(k)$ would not be invertible and would have zero eigenvalues. Next, let $\{ e_1,e_2,\ldots,e_n \}$ be a basis of eigenvectors of $M(k)$ with corresponding eigenvalues $\lambda_1^2,\lambda_2^2,\ldots,\lambda_n^2$. Then, the $2n$ vectors
$$
(e_j,\pm \lambda_j^{-1} R(k) e_j),\qquad j=1,2,\ldots,n
$$
are linearly independent eigenvectors of $P(k)$ with nonzero eigenvalues $\pm\lambda_j$. Furthermore, since $Q(k)$ has full rank,
$$
\dim\ker Q(k) = m - n,
$$
which provides the remaining $m - n$ linearly independent eigenvectors of $P(k)$, which have zero eigenvalues.

To prove (b) we first note that the hypothesis implies that $M(k)$ is diagonalizable with strictly positive eigenvalues, so that the statement in (a) holds. It remains to show that a smooth symmetrizer $H(k)$ can be constructed for $P(k)$. We make the ansatz
\begin{equation}
H(k) = \left( \begin{array}{cc} H_1(k) & 0 \\ 0 & H_2(k) \end{array} \right)
\label{Eq:HkDiag}
\end{equation}
with $H_2(k)$ a symmetric, positive-definite $m\times m$ matrix to be determined. The condition for $H(k) P(k)$ to be symmetric is equivalent to $H_1(k) Q(k) = R(k)^T H_2(k)$. Instead of $H(k)$ we may equip the vector spaces $V$ and $W$ with the scalar products
\begin{eqnarray}
(v,v')_1 &:=& u^T H_1(k) v',\qquad v,v'\in V,\\
(w,w')_2 &:=& w^T H_2(k) w',\qquad w,w'\in W,
\end{eqnarray}
and the condition for $H(k) P(k)$ to be symmetric can be rewritten as
\begin{equation}
(v,Q(k)w)_1 = (R(k) v,w)_2,\qquad v\in V, w\in W,
\label{Eq:QR}
\end{equation}
that is, $R(k)$ is the adjoint of $Q(k)$. In order to define $H_2(k)$ or, equivalently, $(\cdot,\cdot)_2$, we denote by $W_0 := \ker Q(k)$ the kernel of $Q(k)$ (which has dimension $m-n$) and by $W_1\subset W$ the image of $R(k)$ (which has dimension $n$). These subspaces are transversal to each other, since $w\in W_0\cap W_1$ implies that $Q(k)w = 0$ and $w = R(k)v$ for some $v\in V$, which in turn implies that $M(k) v = Q(k) R(k) v = 0$ and hence $v = 0$ and $w = 0$. Therefore, we can decompose
$$
W = W_0 \oplus W_1,
$$
and correspondingly, each $w\in W$ can be written uniquely in the form $w = w_0 + w_1$ with $w_0\in W_0$ and $w_1\in W_1$. Let $\overline{R}(k): V\to W_1$ denote the restriction of $R(k)$ to its image, and $\overline{R}(k)^{-1}: W_1\to V$ its inverse. Then, we define
$$
(w,w')_2 := w_0^T w_0' + (\overline{R}(k)^{-1} w_1)^T H_1(k) M(k) \overline{R}(k)^{-1} w_1'
$$
for $w,w'\in W$, which is clearly symmetric, positive-definite, and smooth in $k$.\footnote{The smoothness property can be established by noticing that the orthogonal projector $\pi_1(k): W\to W$ onto $W_1$ is given by
$$
\pi_1(k) = R(k)\left[ R(k)^T R(k) \right]^{-1} R(k)^T,
$$
which is smooth in $k$, such that $\overline{R}(k)^{-1} w_1 = \left[ R(k)^T R(k) \right]^{-1} R(k)^T w$ for all $w\in W$. Likewise, the orthogonal projector $\pi_2(k)$ onto $W_0$ is given by
$$
\pi_2(k) = I - Q(k)^T\left[ Q(k) Q(k)^T \right]^{-1} Q(k),
$$
which is smooth in $k$.} We now show that this scalar product is such that the relation~(\ref{Eq:QR}) is satisfied. For this, let $v\in V$ and $w = w_0 + w_1\in W$. Then, $R(k) v\in W_1$, and hence $\overline{R}(k)^{-1} R(k)v = v$, and also $R(k)\overline{R}(k)^{-1} w_1 = w_1 = w - w_0$. Therefore,
\begin{eqnarray*}
(R(k)v, w)_2 &=& v^T H_1(k) M(k)\overline{R}(k)^{-1} w_1 \\
 &=& (v, M(k)\overline{R}(k)^{-1} w_1 )_1 \\
 &=& (v, Q(k) w - Q(k)w_0 )_1 = (v,Q(k) w)_1,
\end{eqnarray*}
which concludes the proof of statement (b).
\qed

\proofof{Lemma~\ref{Lem:Auxiliary}} If $\ve{a}$ and $\ve{b}$ are orthogonal to each other, one can introduce an orthonormal basis $\ve{e}_1,\ve{e}_2,\ldots,\ve{e}_n$ of $\Real^n$ such that $\ve{e}_1$ and $\ve{e}_2$ are parallel to $\ve{a}$ and $\ve{b}$, respectively. In this basis,
$$
{\cal M}\ve{e}_1 = \ve{e}_1,\qquad
{\cal M}\ve{e}_2 = \ve{e}_2 + |\ve{a}| |\ve{b}| \ve{e}_1,  
$$
and ${\cal M}\ve{e}_j = \ve{e}_j$ for $j = 3,4,\ldots n$. Therefore, ${\cal M}$ has a non-trivial Jordan block and is not diagonalizable.

From now on, suppose $\ve{a}$ and $\ve{b}$ are not orthogonal to each other. Let $\ve{e}_1$ be a unit vector parallel to $\ve{a}$ and let $\ve{e}_2,\ldots,\ve{e}_n$ be a basis of $\ker(\ve{b}^T)$. Since $\ve{b}^T\ve{a} \neq 0$, the vectors $\ve{e}_1,\ve{e}_2,\ldots,\ve{e}_n$ form a basis of $\Real^n$ with respect to which
$$
{\cal M}\ve{e}_1 = (1 + \ve{b}^T\ve{a})\ve{e}_1,\qquad
{\cal M}\ve{e}_j = \ve{e}_j,\quad j=2,3,\ldots n,
$$
which shows that $M$ is diagonalizable with eigenvalues $1 + \ve{a}^T\ve{b}$ and $1$.

Finally, we note that a given vector $\ve{v}\in \Real^n$ satisfies $(\ve{v},{\cal H}\ve{v}) = 0$ only if $\ve{v}$ is proportional to $\ve{a}$ and orthogonal to $\ve{b}$ at the same time. Since $\ve{a}^T\ve{b}\neq 0$ this is only posible if $\ve{v} = \ve{0}$, which shows that ${\cal H} = {\cal H}^T$ is positive definite. Further, a simple calculation reveals that
$$
{\cal H} {\cal M} = \kappa_0( |\ve{a}|^2 I - \ve{a}\ve{a}^T) + \kappa_1(1 + \ve{a}^T\ve{b})\ve{b}\ve{b}^T,
$$
which is clearly symmetric.
\qed

\bibliography{shortbib}

\end{document}